\documentclass[showpacs]{revtex4-1}
\usepackage{amsmath}
\usepackage{amssymb}
\usepackage{epsfig}

\newcommand{\countzero}{\setcounter{equation}{0}%
         }
\newlength{\feynwidth} 
\newlength{\feynwid} 

\newcommand{\La}{{\Lambda}}
\newcommand{\Si}{{\Sigma}}
\newcommand{\C}{{C_{\xi}}}
\newcommand{\be}{\begin{eqnarray}}
\newcommand{\ee}{\end{eqnarray}}

\begin{document} 
\title{Strangeness $S=-2$ baryon-baryon interaction at next-to-leading order 
in chiral effective field theory}

\author{J. Haidenbauer$^{1,2}$, Ulf-G. Mei\ss ner$^{1,2,3}$, S. Petschauer$^{4}$}

\affiliation{
$^1$Institute for Advanced Simulation, Forschungszentrum J\"ulich, D-52425 J\"ulich, Germany \\
$^2$Institut f\"ur Kernphysik (Theorie) and J\"ulich Center for 
Hadron Physics, Forschungszentrum J\"ulich, D-52425 J\"ulich, Germany \\
$^3$Helmholtz Institut f\"ur Strahlen- und Kernphysik and Bethe Center
 for Theoretical Physics, Universit\"at Bonn, D-53115 Bonn, Germany \\
$^4$Physik Department, Technische Universit\"at M\"unchen, D-85747
  Garching, Germany
}

\begin{abstract}
The strangeness $S=-2$ baryon-baryon interaction is studied in chiral effective field 
theory up to next-to-leading order. The potential at this order consists of contributions 
from one- and two-pseudoscalar-meson exchange diagrams and from four-baryon contact 
terms without and with two derivatives. 
SU(3) flavor symmetry is imposed for constructing the interaction in the $S=-2$ sector. 
Specifically, the couplings of the pseudoscalar mesons to the baryons are fixed by
SU(3) symmetry and, in general, also the contact terms are related via SU(3) symmetry 
to those determined in a previous study of the $S=-1$ hyperon-nucleon interaction.
The explicit SU(3) symmetry breaking due to the physical masses of the pseudoscalar 
mesons ($\pi$, $K$, $\eta$) is taken into account.
It is argued that the $\Xi N$ interaction has to be relatively weak to be in 
accordance with available experimental constraints. In particular, the published 
values and upper bounds for the $\Xi^- p$ elastic
and inelastic cross sections apparently rule out a somewhat stronger attractive 
$\Xi N$ force and, specifically, disfavor any near-threshold deuteron-like 
bound states in that system.
\end{abstract}
\pacs{13.75.-n,13.75.Ev,12.39.Fe,25.80.Pw}

\maketitle

\section{Introduction}

The interaction between baryons in the strangeness $S=-2$ sector 
($\Lambda\Lambda$, $\Xi N$, $\Sigma\Lambda$, $\Sigma\Sigma$) has been
in the focus of interest for many years. 
Here,  the so-called $H$-dibaryon has certainly played 
a prominent role. The $H$-dibaryon, a deeply bound \hbox{6-quark} state with $S=-2$, $J=0$ 
and isospin $I=0$, was predicted by Jaffe within the bag model~\cite{Jaffe:1976yi} 
and should appear in the coupled $\La\La - \Xi N - \Si\Si$ system.
So far none of the experimental searches for the $H$-dibaryon led to
convincing signals \cite{Yoon07,Kim:2013}.
Recently, evidence for a bound $H$-dibaryon was claimed based on
lattice QCD calculations \cite{Beane,Inoue,Beane11a}. Extrapolations
of those computations, performed for $m_\pi \gtrsim 400$~MeV, to the physical pion
mass suggest, however, that most likely the $H$-dibaryon disappears into the 
continuum \cite{Beane11,Shanahan11,JMH1,JMH2,Inoue11a,Shanahan:2013}.

The interaction in the strangeness $S=-2$ sector has also gained 
increasing importance in discussions on the properties of neutron stars in
the context of the problem known as the ``hyperon puzzle'' \cite{Vidana:2015}. 
It concerns the observation of two-solar-mass neutron stars \cite{Demorest2010,Antoniadis2013} 
which provides particularly strong restrictions for the appearance of hyperons 
in neutron stars. The latter leads to a softening of the equation-of-state (EoS) 
and to a reduction of the attainable maximum mass. A repulsive hyperon-hyperon 
($YY$) interaction is seen as one of the 
possible mechanisms that could generate additional repulsion so that 
the EoS remains sufficiently stiff  
\cite{Weissenborn:2011,Oertel:2014,Maslov:2015,Lonardoni:2013,Lim:2014}, 
as is required for the explanation of neutron stars with such high masses.

Another debated issue is the existence of $\Xi$ hypernuclei \cite{Ueda:2001,Gal:2007,Hiyama10}. 
The observed spectrum of the $(K^-,K^+)$ reaction on a $^{12}$C target \cite{Khaustov} 
has been viewed as indication for a moderately attractive $\Xi N$ interaction. In particular, 
an initial analysis of this reaction resulted in an
attractive $\Xi$ potential of $U_\Xi \approx -14$~MeV \cite{Khaustov}. 
On the other hand, a re-analysis by Kohno et al. came to the conclusion 
that an almost zero \cite{Kohno:2010} or even a weakly repulsive $\Xi$ 
potential \cite{Kohno} is preferable.  
Very recently first evidence for the existence of a deeply bound $\Xi^-$-$^{\,14}$N 
system was claimed~\cite{Nakazawa:2015}, that again would be indicative for at least 
some attraction. On the extreme side, there are speculations that the $\Xi N$ interaction 
could be much more strongly attractive~\cite{Garcilazo:2015b} 
so that even $\Xi NN$ bound states could exist.

We present an investigation of the baryon-baryon ($BB$) interaction in the strangeness 
$S=-2$ sector within SU(3) chiral effective field theory (EFT). 
The work is an extension of our previous leading-order study \cite{Polinder07} 
to next-to-leading order (NLO) in the chiral expansion. 
Chiral EFT as a tool to treat the interaction between baryons was initially proposed by 
Weinberg \cite{Wei90,Wei91} for the nucleon-nucleon ($NN$) system and rather successfully
put into practice in a series of works by Epelbaum et al.~\cite{Epe05,Epelbaum:2014} 
and Entem and Machleidt~\cite{Entem:2003ft,Entem:2015}, 
see also the reviews \cite{Epelbaum:2005pn,Machleidt:2011}. 
The application of this scheme to the hyperon-nucleon ($YN$) sector 
\cite{Kor01,Polinder:2006,Haidenbauer13} led to satisfying results as well.
Specifically, as demonstrated in Ref.~\cite{Haidenbauer13},
at NLO in chiral EFT the $\Lambda N$ and $\Sigma N$ scattering data can be
reproduced with the same level of quality as by the most advanced 
meson-exchange $YN$ interactions.
Furthermore, the properties of hyperons in nuclear matter are nicely described 
with such an NLO interaction \cite{Haidenbauer:2014,Petschauer:2015}.
Therefore, it is timely to also provide now a NLO $BB$ interaction for the $S=-2$ 
sector. Nevertheless, it has to be said that the overall situation is not that 
encouraging because of the lack of pertinent scattering data for the $\La\La$,
$\Si\Si$ and $\Xi N $ systems. Indeed, there are just a few data 
points and upper bounds for the $\Xi N$ elastic and inelastic cross sections 
\cite{Ahn:1997,Aoki:1998,Ahn:2006} that put constraints on the corresponding 
interactions. 
Clearly, the use of chiral EFT cannot overcome that problem 
already faced by phenomenological approaches to the $BB$ interaction in the 
$S=-2$ sector \cite{Tominaga:1998,Stoks99,Rijken:2006ep,Fujiwara:2006,Rijken:2010}.
However, once more data will be available, chiral EFT is the premier framework 
to analyze these. 

\begin{figure}[t]
 \centering
 \includegraphics[width=\feynwidth]{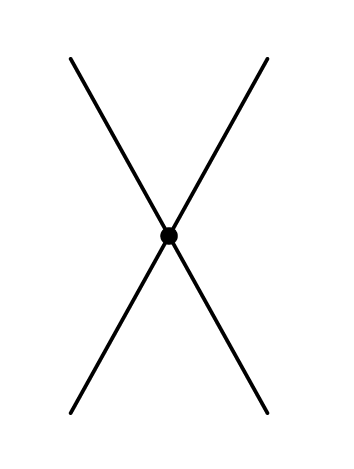}
 \includegraphics[width=\feynwidth]{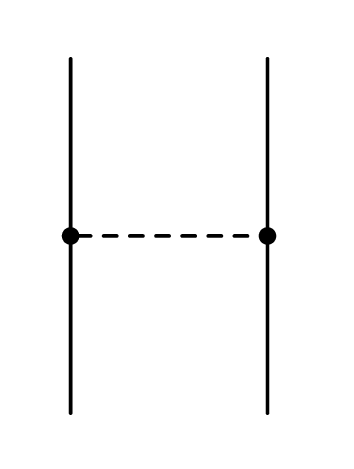}
 \includegraphics[width=\feynwidth]{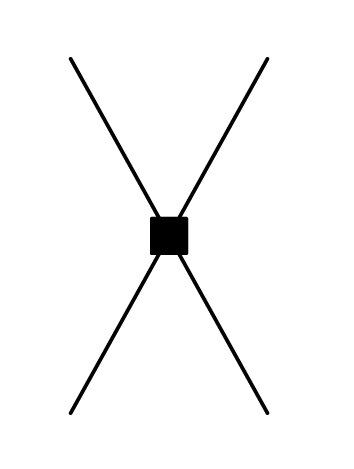}
 \includegraphics[width=\feynwidth]{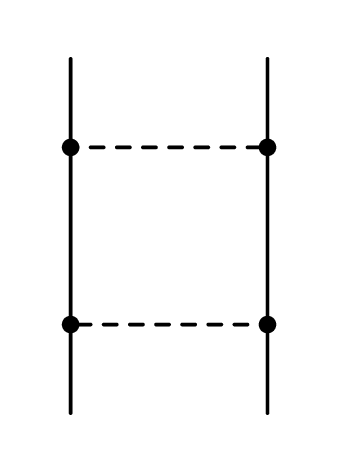}
 \includegraphics[width=\feynwidth]{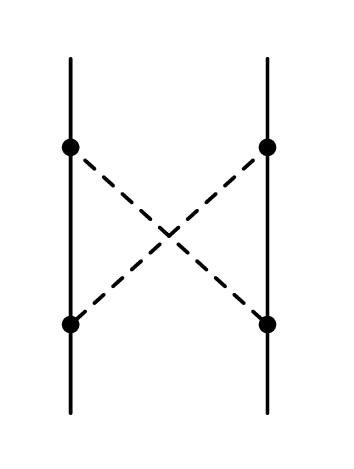}
 \includegraphics[width=\feynwidth]{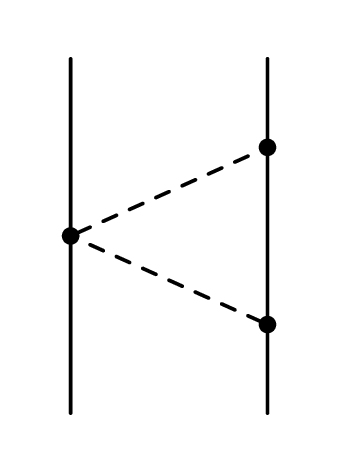}
 \includegraphics[width=\feynwidth]{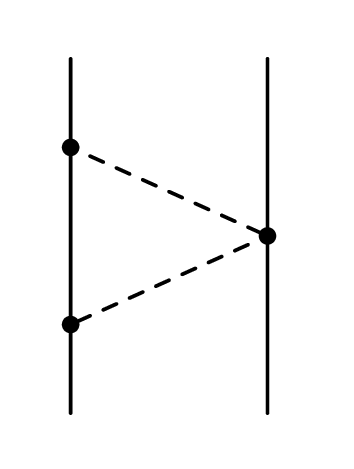}
 \includegraphics[width=\feynwidth]{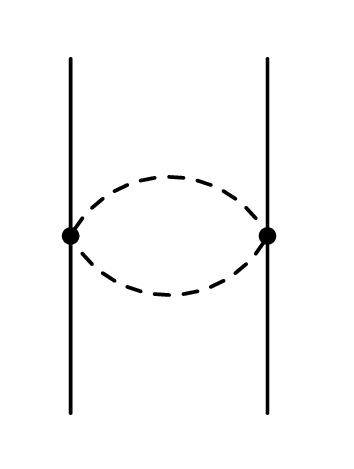}
 \caption{Relevant Feynman diagrams up-to-and-including next-to-leading order. Solid and 
dashed lines denote octet baryons and pseudoscalar mesons, respectively. The square 
symbolizes a contact vertex with two derivatives.
From left to right: LO contact term, one-meson exchange, NLO contact term, planar box, 
crossed box, left triangle, right triangle, football diagram.
From the planar box graph, only the irreducible part contributes to the potential.
} 
\label{fig:feynman}
\end{figure}

We begin with providing an overview of the available experimental information and 
constraints on the $\Lambda\Lambda$ and $\Xi N$ interactions (in Sect.~II), and then 
proceed with a brief introduction to the concepts and formalism of chiral EFT. 
This approach exploits the symmetries of QCD together with the appropriate 
low-energy degrees of freedom to construct the $BB$ interaction.
Essential features of chiral EFT are that the results can be improved systematically 
by going to higher order in the power counting scheme, and that two- and three-baryon 
forces can be calculated in a consistent way \cite{Petschauer15}.
At the order we are working the chiral interaction consists of $BB$ contact terms
without derivatives and with two derivatives, together with contributions from
one-pseudoscalar-meson exchanges and from (irreducible) two-pseudoscalar-meson exchanges 
\cite{Haidenbauer13}.
The corresponding diagrams are shown schematically in Fig.~\ref{fig:feynman}.
The contributions from pseudoscalar-meson exchanges to the $BB$ interaction are 
completely fixed by the assumed SU(3) flavor symmetry. On the other hand, the strength 
parameters associated with the contact terms, the low-energy constants (LECs), need to 
be determined in a fit to data. How this is done is discussed in Sect.~III in detail. 
We impose SU(3) symmetry also for the contact terms which reduces the number of 
independent LECs that can contribute significantly. Furthermore, in general, we 
exploit SU(3) symmetry to relate the LECs needed in the $S=-2$ sector to 
those fixed in our fit to the $S=-1$ $Y N$ data before \cite{Haidenbauer13}.

Results for $\Lambda\Lambda$ and $\Xi N$ scattering are presented in Sect.~IV. 
As the main outcome, those indicate that the $\Xi N$ interaction has to be relatively
weak if one takes the available experimental information and, in particular,
the published upper bounds for cross sections into consideration.
A weak $\Xi N$ interaction has been already predicted by our leading-order (LO) 
interaction \cite{Polinder07} and also by several phenomenological models of the
$BB$ interaction in the $S=-2$ sector \cite{Stoks99,Fujiwara:2006,Gasparyan2011}.

The paper closes with a Summary and with an Appendix that contains some
technical details. 

\section{Available experimental information}

Before we come to our calculation let us briefly review the experimental
situation for the strangeness $S=-2$ sector. Unfortunately, there are practically 
no genuine scattering data for the $\Lambda\Lambda$ and $\Xi N$ systems 
at low energies. 
However, there is other information from few- and many-baryon systems that
allows one to put at least some qualitative constraints on the 
baryon-baryon interaction with strangeness $S=-2$. Those constraints 
are used as guideline in the construction of our interaction. 

\subsection{The $\Lambda\Lambda$ system}

To our knowledge there are no scattering data for $\Lambda\Lambda$. However, 
available experimental information on double-$\Lambda$ hypernuclei put fairly tight 
constraints on the $\Lambda\Lambda$ $^1S_0$ scattering length. Specifically, 
the unambiguously identified ${}_{\Lambda\Lambda}^{\;\;\;6}{\rm He}$
hypernucleus (Nagara event) \cite{Takahashi:2001nm} plays an important
role in this context. The analysis of that event yielded the fairly small 
separation energy of $\Delta B_{\La\La} = B_{\La\La} ( _{\La\La}^{\phantom{6}6}{\rm He})
- 2 B_{\La} ( _{\La}^5{\rm He}) = 1.01\pm0.20$~MeV. After a re-evaluation in 
2010 the separation energy is now given as 
$\Delta B_{\Lambda\Lambda} = 0.67\pm0.17$~MeV \cite{Nakazawa}. 
Calculations of the ${}_{\Lambda\Lambda}^{\;\;\;6}{\rm He}$ hypernucleus have been 
performed in a variety of different approaches such as three-body Faddeev equations 
applied to the cluster model, in the Brueckner theory approach, or with the stochastical 
variational method. Those suggest that the $\Lambda\Lambda$ scattering length should 
be of the order of $-1.3$ to $-0.5$~fm 
\cite{Filikhin02,Filikhin04,Afnan03,Rijken:2006ep,Fujiwara:2006,Yamada:2004,Vidana:2004,Usmani:2004,Nemura05}
to obtain separation energies in line with the Nagara event.

Information on the scattering length of two-baryon systems like $\Lambda\Lambda$ 
can be also extracted from corresponding final-state interactions in production reactions. 
A method based on dispersion theory that also allows for a quantitative estimation of the 
theoretical uncertainties has been proposed in Ref.~\cite{Gasparyan2003}. 
Its application to available data on the $\Lambda\Lambda$ invariant mass
from the reaction $^{12}C(K^-,K^+\Lambda\Lambda X)$ \cite{Yoon07} yielded 
a $^1S_0$ scattering length of $a=-1.2\pm 0.6$~fm \cite{Gasparyan2011}. 

Finally, there are constraints for the $\Lambda\Lambda$ scattering length
emerging from studies of $\Lambda\Lambda$ correlations in relativistic 
heavy-ion collisions. In particular, the analysis of Morita et al.~\cite{Morita} 
of recent data by the STAR collaboration \cite{Adamczyk} suggests 
values of $-1.25 < a < -0.56$~fm. It is somewhat disturbing, however, that the 
experimentalists themselves obtained the rather different value of 
$a = 1.10 \pm 0.37^{+0.08}_{-0.68}$~fm in their own analysis \cite{Adamczyk}. 
Note that a different sign convention is used in that publication. 

\subsection{The $\Xi N$ system}

Results of a measurement of $\Xi N$ scattering at low energies have been
published by Ahn et al.~\cite{Ahn:2006}, where upper limits for the $\Xi^-p$
elastic cross section and for $\Xi^-p\to\Lambda\Lambda$ are given 
for the $p_{\Xi^-}$ momentum range of $0.2$~GeV/c to $0.8$~GeV/c.
Those limits, $24$~mb for the former and $12$~mb for the latter, 
are based on the observation of one event and a null result, respectively.
The cross section for $\Xi^-p\to\Lambda\Lambda$ was also evaluated
from three events observed in the reaction $^{12}C(\Xi^-,\Lambda\Lambda) X$
assuming quasi-free scattering and found to be $4.3^{+6.3}_{-2.7}$~mb. 
A cross section of $4.19\pm 1.9$~mb was reported recently by the same
group, based on a larger number of events \cite{Kim:2015S}. 
In an earlier experiment \cite{Ahn:1997} the same collaboration had already 
deduced the total cross section for $\Xi^-p$ inelastic scattering 
to be around $13$~mb (from nuclear emulsion) and $15$~mb (from the 
$\Xi^-$ absorption probability for $^{12}$C) at $p_{\Xi^-} \simeq 0.6$~GeV/c. 
Since inelastic scattering involves the reactions 
$\Xi^-p\to\Lambda\Lambda$ and $\Xi^-p\to\Xi^0 n$, the cross-section measurement
for the former process mentioned above suggests that $\sigma_{\Xi^-p\to\Xi^0 n}$ 
is of the order of $10$~mb \cite{Ahn:2006}. 

In Ref.~\cite{Aoki:1998} Aoki et al. reported in-medium results. 
The inelastic cross section for $\Xi^-N$ in a nucleus was deduced
to be $12.7^{+3.5}_{-3.1}$~mb at $p_{\Xi^-} = 0.4-0.6$~GeV/c, 
from a measurement of the quasifree $p(K^-,K^+)\Xi^-$ reaction in 
nuclear emulsion.
The cross section obtained from the number of secondary interactions
of $\Xi^-$ particles in the emulsion was estimated to be $20\pm 7$~mb. 
In-medium results have been also presented by Tamagawa et al.~\cite{Tamagawa}. 
In this publication a $\Xi^-N$ elastic cross section of $30.7\pm 6.7^{+3.7}_{-3.6}$~mb 
for an average $\Xi$ momentum of $p_\Xi = 550$~MeV/c is reported. Furthermore, the 
ratio of the $\Xi^-p$ and $\Xi^-n$ scattering cross section was deduced
to be $1.1^{+1.4}_{-0.7}\,^{+0.7}_{-0.4}$. 

There are also results for $\Xi N$ cross sections for various
elastic and inelastic channels at $\Xi$ momenta above $1$~GeV/c
\cite{Charlton,Dalmeijer,Muller,Hauptman} from analyses of 
bubble chamber images. Such momenta are too high to be considered in 
the present study. However, those data can still serve as qualitative 
guideline for the trend of the $\Xi N$ cross sections with increasing energy. 
In particular, one does not expect the cross sections around $0.6 - 1$~GeV/c, 
say, to be significantly larger than the ones observed in those experiments. 
 
Experimental information on the $\Xi N$ invariant mass spectrum
is scarce \cite{Goyal80,Saclay82,Godbersen95}. 
In an measurement performed at 
Saclay~\cite{Saclay82} the missing mass ($MM$) in the reaction 
$K^- + d \to K^+ + MM$ at 1.4~GeV/c was studied. The curve 
presented in this publication exhibits a rather smooth behaviour
around the $\Xi N$ threshold which suggests that the $\Xi^- n$
interaction in the $^1S_0$ and/or $^3S_1$ might be fairly
weak. Such a conjecture is actually in line with the results 
of the majority of the potential models which predict rather 
small $\Xi^- n$ scattering lengths, see Table~IV
of Ref.~\cite{Gasparyan2011} for an overview. 
Also the invariant-mass spectrum published in Ref.~\cite{Goyal80} for the
$\Xi^- p$ case, obtained from a $K^-d$ bubble-chamber experiment,
does not show any enhancement near the $\Xi N$ threshold. However, 
the statistics is too low for drawing any solid conclusions. 

As already mentioned in the Introduction, an indication for an at least 
moderately attractive $\Xi N$ interaction has been seen in the observed 
spectrum of the $(K^-,K^+)$ reaction on a $^{12}$C 
target \cite{Khaustov}. Here an initial theoretical analysis pointed to an
attractive $\Xi$ single particle (s.p.) potential of $U_\Xi(p_\Xi=0) \approx -14$~MeV. 
However, in a re-analysis Kohno et al.~\cite{Kohno:2010}
came to the conclusion that an almost zero s.p. potential is preferable.  
Indeed, the in-medium calculation reported in Ref.~\cite{Kohno}, based on our 
LO $\Xi N$ interaction \cite{Polinder07}, yields $U_\Xi(0)\approx +6$~MeV at nuclear 
matter saturation density ($k_F=1.35$~fm$^{-1}$) and could be still compatible 
with the experiment, judging from the various curves presented in \cite{Kohno:2010}.
See also the related work \cite{Quark:2010} by the same authors. 
Another facet to this somewhat controversal situation was added by 
the first evidence for the existence of a deeply bound $\Xi^-$-$^{\,14}$N 
system reported in Ref.~\cite{Nakazawa:2015}.

\section{Chiral potential at next-to-leading order}

\subsection{General structure}
The chiral interaction at NLO consists of $BB$ contact terms without derivatives 
and with two derivatives, together with contributions from
one-pseudoscalar-meson exchanges and from (irreducible) two-pseudoscalar-meson exchanges
\cite{Haidenbauer13}.
The derivation of the baryon-baryon potentials for strangeness in SU(3) chiral EFT 
up to NLO is described in detail in Refs.~\cite{Haidenbauer13,Petschauer13}.
Specifically, the contact terms are discussed in Sect.~2 of 
Ref.~\cite{Haidenbauer13} (see also Ref.~\cite{Petschauer13}) together with
the one-pseudoscalar-meson exchange potential, while the various contributions
from two-pseudoscalar-meson exchange are documented in Appendix~A of 
Ref.~\cite{Haidenbauer13}. Since the spin-momentum structure is the same for the 
$YN$ and $YY$ systems we do not reproduce the pertinent expressions here. The only
difference occurs in the SU(3) (and isospin) structure and, therefore, we
focus just on this aspect. With regard to the contact terms the potentials
in the various channels are again of the generic form \cite{Haidenbauer13}
\begin{equation}
V=\tilde C+C\,(p^2+p'^2) \ ({\rm for } \ S \ {\rm waves}), \ \ V=C\,pp' 
 \ ({\rm for } \ P \ {\rm waves}),
\label{eq:LEC}
\end{equation}
etc., but the $C$'s are now given by different combinations of the basic set of LECs 
that correspond to the irreducible SU(3) representations  
$8\otimes 8 = 27 \oplus 10^* \oplus 10 \oplus 8_s \oplus 8_a \oplus 1$, 
relevant for scattering of two octet baryons \cite{Swa63,Dover1991}.
The pertinent relations are summarized in Table~\ref{tab:SU3}. For convenience and 
orientation those for the strangeness $S=0$ ($NN$) and $S=-1$ ($\Lambda N$, 
$\Sigma N$) are listed, too. The quantities $p$ and $p'$ in Eq.~(\ref{eq:LEC})
denote the center-of-mass momenta in the initial and final $BB$ states. 

\begin{table}[ht]
\caption{SU(3) relations for the various contact potentials in the isospin basis.
$C^{27}_{\xi}$ etc. refers to the corresponding irreducible SU(3) representation
for a particular partial wave ${\xi}$. The isospin is denoted by $I$. 
The actual potential still needs to be multiplied by pertinent powers of the 
momenta $p$ and $p'$, see Eq.~(\ref{eq:LEC}). 
}
\label{tab:SU3}
\vskip 0.1cm
\renewcommand{\arraystretch}{1.2}
\centering
\begin{tabular}{|l|c|c|c|c|c|c|}
\hline
&Channel &I &\multicolumn{4}{|c|}{$V({\xi})$} \\
\hline
\hline
&        &  &$\xi= \, ^1S_0, \, ^3P_0, \, ^3P_1, \, ^3P_2 $
& $\xi = \, ^3S_1, \, ^3S_1$-$^3D_1, \, ^1P_1$ & $\xi = \, ^1P_1$$\to$$^3P_1$ & $\xi = \, ^3P_1$$\to$$^1P_1$  \\
\hline
${S=\phantom{-}0}$&$NN\rightarrow NN$ &$0$ & \ \ -- & $\C^{10^*}$ &  -- &  --\\
                       &$NN\rightarrow NN$ &$1$ & $\C^{27}$ &  -- &  -- &  --\\
\hline
\hline
${S=-1}$&$\La N \rightarrow \La N$ &$\frac{1}{2}$ &$\frac{1}{10}\left(9C^{27}_{\xi}+C^{8_s}_{\xi}\right)$
& $\frac{1}{2}\left(C^{8_a}_{\xi}+C^{10^*}_{\xi}\right)$ & ${-}C^{8_{sa}}_{\xi}$&  ${-}C^{8_{sa}}_{\xi}$ \\
&$\La N \rightarrow \Si N$ &$\frac{1}{2}$        &$\frac{3}{10}\left(-C^{27}_{\xi}+C^{8_s}_{\xi}\right)$
& $\frac{1}{2}\left(-C^{8_a}_{\xi}+C^{10^*}_{\xi}\right)$ & ${-3} C^{8_{sa}}_{\xi}$& $ C^{8_{sa}}_{\xi}$ \\
&$\Si N \rightarrow \Si N$  &$\frac{1}{2}$        &$\frac{1}{10}\left(C^{27}_{\xi}+9C^{8_s}_{\xi}\right)$
& $\frac{1}{2}\left(C^{8_a}_{\xi}+C^{10^*}_{\xi}\right)$ & ${3} C^{8_{sa}}_{\xi} $ & ${3} C^{8_{sa}}_{\xi}$\\
&$\Si N \rightarrow \Si N$  &$\frac{3}{2}$        &$C^{27}_{\xi}$
& $C^{10}_{\xi}$ & \ \ -- & \ \ -- \\
\hline
\hline
${S=-2}$&$\La\La \rightarrow \La\La$ &$0$   &$\frac{1}{40}\left(27C^{27}_{\xi}+8C^{8_s}_{\xi}+5{C^{1}_{\xi}}\right)$ & --& -- & --\\
&$\La\La\rightarrow \Xi N$ &$0$  &$\frac{-1}{40}\left(18C^{27}_{\xi}-8C^{8_s}_{\xi}-10{C^{1}_{\xi}}\right)$ & --& -- & $2C^{8_{sa}}_{\xi}$ \\ 
&$\La\La\rightarrow \Si\Si$ &$0$ &$\frac{\sqrt{3}}{40}\left(-3C^{27}_{\xi}+8C^{8_s}_{\xi}-5{C^{1}_{\xi}}\right)$ & --& --& --\\
&$\Xi N \rightarrow \Xi N$ &$0$  &$\frac{1}{40}\left(12C^{27}_{\xi}+8C^{8_s}_{\xi}+20{C^{1}_{\xi}}\right)$ & $C^{8_a}_{\xi}$ &$2C^{8_{sa}}_{\xi}$ & $2C^{8_{sa}}_{\xi}$ \\
&$\Xi N \rightarrow \Si\Si$ &$0$ &$\frac{\sqrt{3}}{40}\left(2C^{27}_{\xi}+8C^{8_s}_{\xi}-10{C^{1}_{\xi}}\right)$ & -- & $2\sqrt{3}C^{8_{sa}}_{\xi}$ & --\\
&$\Si\Si\rightarrow \Si\Si$ &$0$ &$\frac{1}{40}\left(C^{27}_{\xi}+24C^{8_s}_{\xi}+15{C^{1}_{\xi}}\right)$ & --& --& --\\
&$\Xi N \rightarrow \Xi N$ &$1$  &$\frac{1}{5}\left(2C^{27}_{\xi}+3C^{8_s}_{\xi}\right)$ & $\frac{1}{3}\left(C^{10}_{\xi}+C^{10^*}_{\xi}+C^{8_a}_{\xi}\right)$ &$-2C^{8_{sa}}_{\xi}$ & $-2C^{8_{sa}}_{\xi}$ \\
&$\Xi N \rightarrow \Si\La$ &$1$ &$\frac{\sqrt{6}}{5}\left(C^{27}_{\xi}-C^{8_s}_{\xi}\right)$ & $\frac{\sqrt{6}}{6}\left(C^{10}_{\xi}-C^{10^*}_{\xi}\right)$ & $\sqrt{\frac{8}{3}}C^{8_{sa}}_{\xi}$ & -- \\
&$\Xi N \rightarrow \Si\Si$ &$1$ & --& $\frac{\sqrt{2}}{6}\left(C^{10}_{\xi}+C^{10^*}_{\xi}-2C^{8_a}_{\xi}\right)$ & -- & $2\sqrt{2}C^{8_{sa}}_{\xi}$ \\
&$\Si\La\rightarrow \Si\La$ &$1$ &$\frac{1}{5}\left(3C^{27}_{\xi}+2C^{8_s}_{\xi}\right)$ & $\frac{1}{2}\left(C^{10}_{\xi}+C^{10^*}_{\xi}\right)$ & -- & --\\
&$\Si\La\rightarrow \Si\Si$ &$1$ & --&$\frac{\sqrt{3}}{6}\left(C^{10}_{\xi}-C^{10^*}_{\xi}\right)$ & -- & $\frac{4}{\sqrt{3}}C^{8_{sa}}_{\xi}$ \\
&$\Si\Si\rightarrow \Si\Si$ &$1$ & --&$\frac{1}{6}\left(C^{10}_{\xi}+C^{10^*}_{\xi}+4C^{8_a}_{\xi}\right)$ & -- & --\\
&$\Si\Si\rightarrow \Si\Si$ &$2$   &$C^{27}_{\xi}$ & --& -- & --\\
\hline
\hline
\end{tabular}
\renewcommand{\arraystretch}{1.0}
\end{table}

The contributions from meson exchange result from taking the appropriate spin-momentum
structure given in Ref.~\cite{Haidenbauer13}, and multiplying it with the corresponding
combination of baryon-baryon-meson coupling constants and with the isospin factors. 
Under the assumption that the coupling constants fulfill the standard SU(3) relations
the contributions from one- and two-meson exchanges do not involve any free parameters. 
The relations between the coupling constants, expressed in terms of the 
axial coupling constant $g_A$, the pion decay constant $f_\pi$, and the so-called 
$F/(F+D)$-ratio, can be found in Appendix~A. Furthermore, we provide there the isospin 
factors for all single and two-meson exchange diagrams that arise in the strangeness 
$S=-2$ sector. 
See also Refs.~\cite{Polinder07,Haidenbauer13,Petschauer13,HMP:2015}. 

Once the potentials are established, a partial-wave projection of them is performed, 
as described in detail in Ref.~\cite{Polinder:2006}. 
The reaction amplitudes are obtained from the solution of a coupled-channel
Lippmann-Schwinger (LS) equation:
\begin{eqnarray}
&&T^{\rho''\rho',J}_{\nu''\nu'}(p'',p';\sqrt{s})=V^{\rho''\rho',J}_{\nu''\nu'}(p'',p')+
\sum_{\rho,\nu}\int_0^\infty \frac{dpp^2}{(2\pi)^3} \, V^{\rho''\rho\, ,J}_{\nu''\nu}(p'',p)
\frac{2\mu_{\nu}}{q_{\nu}^2-p^2+i\eta}T^{\rho\rho',J}_{\nu\nu'}(p,p';\sqrt{s})\ .
\label{LS} 
\end{eqnarray}
Here,
the label $\nu$ indicates the particle channels and the label $\rho$ the partial wave. $\mu_\nu$
is the pertinent reduced baryon mass. The on-shell momentum $q_{\nu}$ in the intermediate state, is
determined by $\sqrt{s}=\sqrt{M^2_{B_{1,\nu}}+q_{\nu}^2}+\sqrt{M^2_{B_{2,\nu}}+q_{\nu}^2}$.
Relativistic kinematics is used for relating the laboratory momentum $p_{{\rm lab}}$ of the hyperons
to the center-of-mass momentum.

We solve the LS equation in the particle basis, in order to incorporate the correct physical
thresholds. This is important for the channels with total charge zero where 
the relatively large mass difference of around $7$~MeV between the $\Xi^0$ and $\Xi^-$ 
\cite{PDG} causes a corresponding splitting between the $\Xi^0 n$ and $\Xi^- p$ 
thresholds.
Depending on the total charge, up to six baryon-baryon channels can couple.
The Coulomb interaction is not taken into account in the present study because there
are simply no near-threshold data that would require such a more elaborate treatment.
The potentials in the LS equation are cut off with a regulator function, $f_R(\Lambda) =
\exp\left[-\left(p'^4+p^4\right)/\Lambda^4\right]$,
in order to remove high-energy components \cite{Epe05}.
We consider cut-off values of $\Lambda=500$ -- $650\,$MeV, i.e. in that range where the best 
results were achieved in our study of the $\La N$ and $\Si N$ interactions \cite{Haidenbauer13}. 
Accordingly, we display our results as bands which represent their variation 
with the cut-off. 
Note that similar $\Lambda$ values are also used for chiral $NN$ potentials \cite{Epe05}.
As discussed later, this cut-off variations gives only a rough estimate of the theoretical 
uncertainty.

\begin{figure}[t]
\includegraphics*[width=5cm]{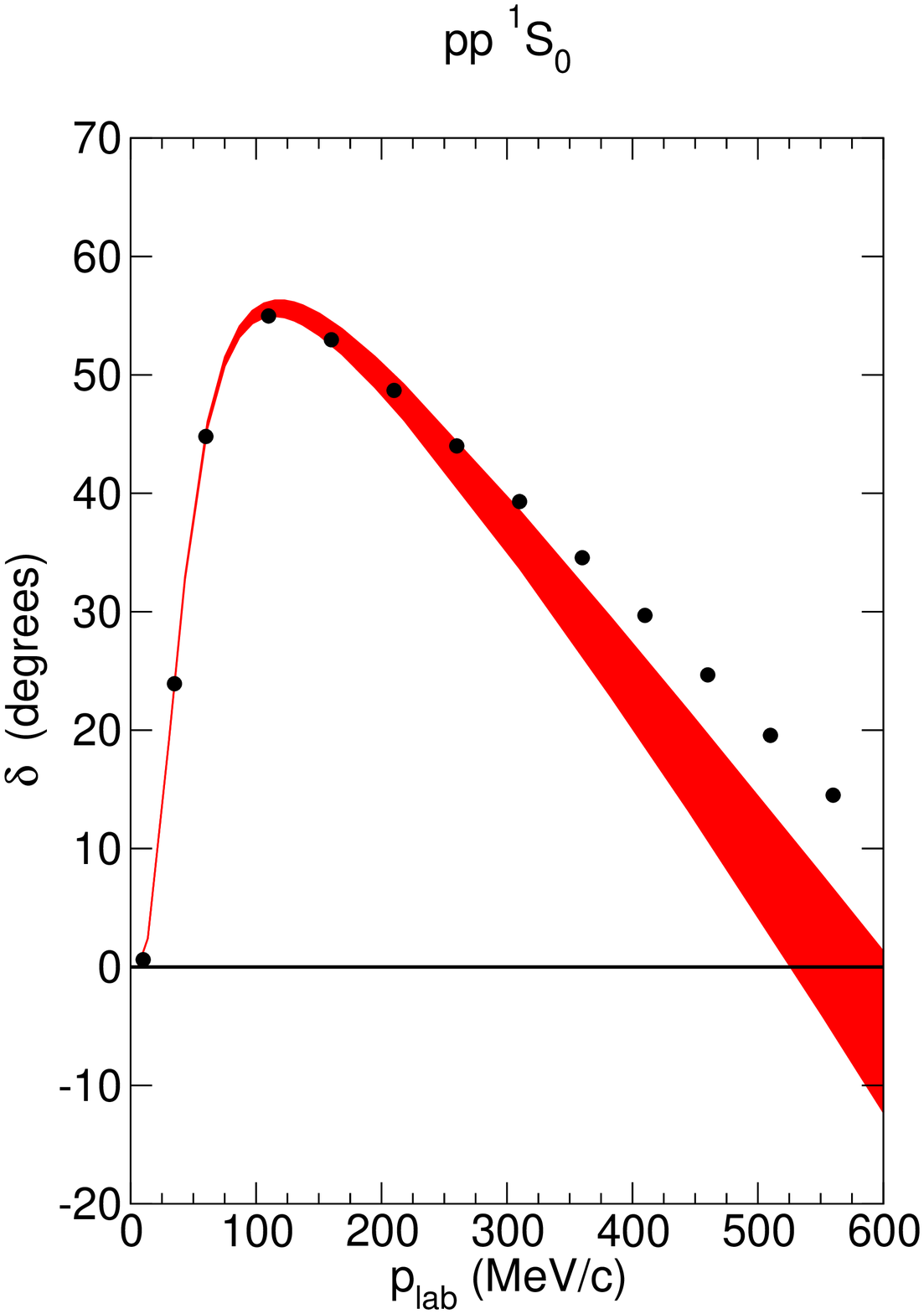}\includegraphics*[width=5cm]{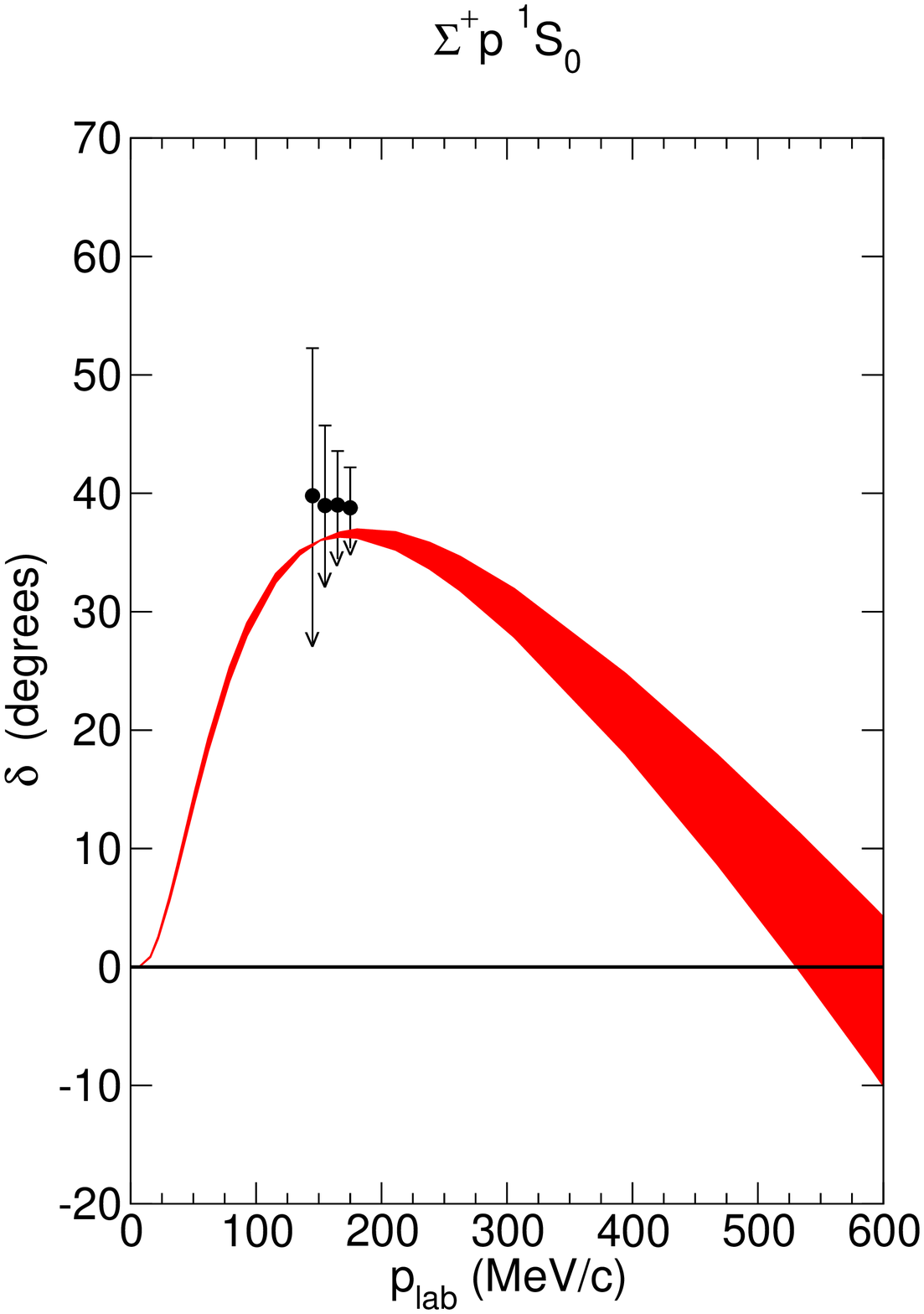}
\includegraphics*[width=5cm]{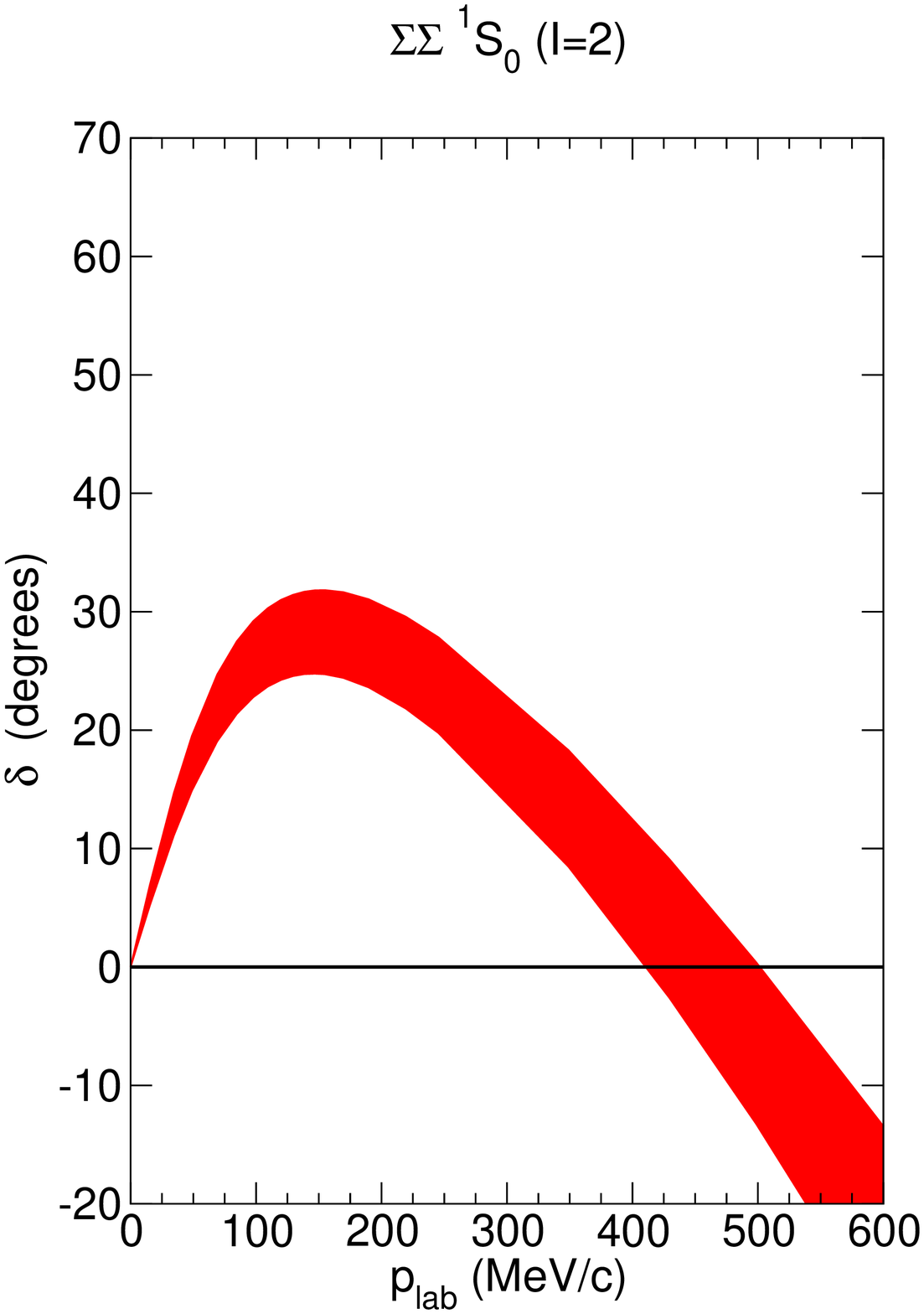}
\caption{$pp$, $\Sigma^+p$, and $\Sigma^+\Sigma^+$ phase shifts in the $^1S_0$ partial wave. 
The filled band represent our results at NLO. 
The $pp$ phase shifts of the GWU analysis \cite{SAID1} are shown by circles.
In case of $\Sigma^+ p$ the circles indicate upper limits for the phase shifts,
deduced from the $\Sigma^+ p$ cross section \cite{HMP:2015}.
The results are taken over from Ref.~\cite{HMP:2015}.
}
\label{NNphase}
\end{figure}

\subsection{Determination of contact terms}

In our EFT study of the $Y N$ interaction \cite{Haidenbauer13} we were able to determine
practically all contact terms (at least for the $S$ waves) by a direct fit to
the pertinent $\La N$ and $\Si N$ data, without any recourse to the $NN$ system. This is not 
possible for the $S=-2$ sector due to the already mentioned limited experimental 
information. Here, in general, we will and have to take over the LECs fixed in 
our fit to the $Y N$ data before. In principle, this is required by a rigorous 
implementation of SU(3) symmetry anyway. 

Having said this, we want to recall that in Ref.~\cite{Haidenbauer13} it had turned out 
that a simultaneous description of the $YN$ data and the $NN$ phase shifts is not possible 
if one really insists on consistent and strictly SU(3) symmetric contact terms. In 
particular, the value of the contact 
term $\tilde C^{27}_{^1S_0}$, required for reproducing the $pp$ (or $np$) $^1S_0$ phase 
shifts on the one hand and for describing the $\Sigma^+ p$ cross section on the
other hand could not be reconciled with the requirements of complete SU(3) symmetry 
given by the relations in Table~\ref{tab:SU3}. 
In the light of this observation, in the present study we will also depart from the 
strict SU(3) connection with the $YN$ sector in cases where we see a strong indication 
for that from the comparison of our results with the $S=-2$ data. 
In practice this concerns the LECs $\tilde C^{27}_{^1S_0}$ and $\tilde C^{8_a}_{^3S_1}$.
 
Before we discuss the details, let us emphasize that this policy is consistent 
with the employed power counting scheme. At NLO SU(3) symmetry breaking in the 
leading $S$-wave contact terms arises already naturally in the perturbative expansion of the 
baryon-baryon potential due to meson mass insertions \cite{Petschauer13}.
Actually, this aspect was exploited by us in a recent study \cite{HMP:2015}. For the $^1S_0$ 
partial wave and for $BB$ channels with maximal isospin
there is only one independent SU(3) symmetry breaking LEC 
for the $S=0$, $-1$, and $-2$ systems, i.e. 
\begin{eqnarray}
\nonumber
V^{(I=1)}_{N N} &=& \tilde C^{27}_{^1S_0} + C^{27}_{^1S_0}(p^2+p'^2)
+\frac{1}{2}C_1^\chi (m^2_K-m^2_\pi) ,\\
\nonumber
V^{(I=3/2)}_{\Sigma N } &=& \tilde C^{27}_{^1S_0} + C^{27}_{^1S_0}(p^2+p'^2)
+\frac{1}{4}C_1^\chi (m^2_K-m^2_\pi) ,\\
V^{(I=2)}_{\Sigma \Sigma } &=& \tilde C^{27}_{^1S_0} + C^{27}_{^1S_0}(p^2+p'^2) , 
\label{LEC}
\end{eqnarray}
in the notation used in Ref.~\cite{HMP:2015}.
The aforementioned $pp$ phase shifts and the $\Sigma^+ p$ cross section can be used to
pin down the symmetry breaking LEC $C_1^\chi$ together with $\tilde C^{27}_{^1S_0}$ 
and $C^{27}_{^1S_0}$ and, therefore, we could give predictions for the $\Sigma\Sigma$ 
interaction with isospin $I=2$. 
The corresponding phase shifts are shown in Fig.~\ref{NNphase}.

In the present study we take over the LECs $\tilde C^{27}_{^1S_0}$ and $C^{27}_{^1S_0}$ 
for the $\Si\Si$ channel as given in Table~3 of Ref.~\cite{HMP:2015} and use them for all 
interactions 
in the $^1S_0$ partial wave in the $S=-2$ sector. The negative value of the SU(3) breaking 
term $C_1^\chi$ \cite{HMP:2015} implies that there is a decrease in the attraction when one 
goes from the $NN$ system to $S=-1$ and then to $S=-2$, see Eq.~(\ref{LEC}). Accordingly, 
the $\Lambda\Lambda$ interaction based on $\tilde C^{27}_{^1S_0}$ and $C^{27}_{^1S_0}$ 
from Ref.~\cite{HMP:2015} is less attractive than if we would have used those determined 
in a fit to the $YN$ data \cite{Haidenbauer13} following strictly SU(3) arguments. 
For the LECs $\tilde C^{8_s}_{^1S_0}$ 
and $C^{8_s}_{^1S_0}$ we adopt the values from our $YN$ study \cite{Haidenbauer13}. 
The interactions in the $^1S_0$ partial wave in the $I=0$ channels involve
also the LECs $\tilde C^1$ and $C^1$ which correspond to the $1$ irreducible
representation, see Table~\ref{tab:SU3}. Those terms do not contribute to the 
$NN$ and $YN$ systems. In an attempt to assign values to those LECs we followed
our previous study \cite{Polinder07} and varied them within limits set roughly
by the so-called natural value which is equal to $4\pi/f_{\pi}^2$ 
for the LO partial-wave projected contact term \cite{Epelbaum:2005pn}.
Since the $\Lambda\Lambda$ interaction is expected to be only moderately attractive, 
as discussed above, we considered only such variations of the $C^1$'s that led to
$\Lambda\Lambda$ scattering lengths in the range of $-1$ to $-0.5$~fm. 
In particular, we excluded regions which resulted in bound states or near-threshold 
resonances in the $\Lambda\Lambda$ or $\Xi N$ systems, for which there is no 
experimental support at present.  

The LECs in the $^3S_1$-$^3D_1$ partial wave are taken over from the $YN$ sector 
\cite{Haidenbauer13}. For that partial wave the symmetry breaking LECs 
are strongly interrelated \cite{Petschauer13} and it is impossible to determine 
any of these by considering the $NN$ phase shifts together with the experimential 
information on the $\Lambda N$ and $\Sigma N$ systems. 
It turns out that the $S=-2$ $BB$ interaction based on those LECs leads to a 
$\Xi^-p$ cross section that exceeds the upper limit given in Ref.~\cite{Ahn:2006}. 
The main reason for that is the appearance of a bound state 
close to the $\Xi N$ threshold in the $^3S_1$-$^3D_1$ partial wave with $I=0$. 
Since only LECs in the $8_a$ representation contribute to this channel, see 
Table~\ref{tab:SU3}, we simply readjust $\tilde C^{8_a}_{^3S_1}$ in order to obtain
an interaction that yields $\Xi N$ results consistent with the experimental bounds. 
Actually, an inspection of Table~3 in \cite{Haidenbauer13} reveals
that the values of $\tilde C^{8_a}_{^3S_1}$ are much much smaller than any of the
other LECs. Thus, one could speculate that those values are not so well determined in
the fit to the $YN$ data anyway. In practice only a small change in $\tilde C^{8_a}_{^3S_1}$
is needed to remove the bound state and to achieve $\Xi N$ cross sections in line 
with the bounds suggested by the experiments. 
 
The contact terms in the $P$ waves are all taken over from Ref.~\cite{Haidenbauer13},
with one exception: 
We readjusted the $C^{8_a}_{^1P_1}$ of the $YN$ interaction of \cite{Haidenbauer13}
so that now the $\Lambda p$ ${^1P_1}$ phase shift is overall repulsive. This has no influence
on the cross section results presented in Ref.~\cite{Haidenbauer13}. However, the
predicted $\Xi N$ cross sections are then smaller at higher momenta and, therefore,
better in line with the behavior suggested by the experiments. 
The values for the $C^1$'s (ocurring in the $^3P_0$, $^3P_1$, and $^3P_2$ partial waves)
have been likewise fixed by the requirement that the $\Xi N$ cross sections should 
remain small for higher energies. 
We do not consider $^1P_1$--$^3P_1$ mixing at this stage, i.e. we put $C^{8_{sa}}$ 
to zero. All low-energy constants used in the present study are tabulated in
Appendix~A. 

While the above strategy allows us to fix all LECs we are certainly far from being 
able to determine them uniquely, as should have become obvious from the preceding discussion.
This is a consequence of the limited experimental information. 
The situation differs from that in our EFT study of the $Y N$ interaction 
where, as noted, practically all contact terms (at least for the $S$ waves) could be 
really fixed by a direct fit to $\Lambda N$ and $\Sigma N$ data.
In view of these circumstances it should be clear that the present investigation on the 
$BB$ interaction in the $S=-2$ sector, performed in chiral EFT up to NLO, can only have 
a preliminary and exploratory status. 

\section{Results}

\begin{figure}
\includegraphics*[width=6cm,angle=-90]{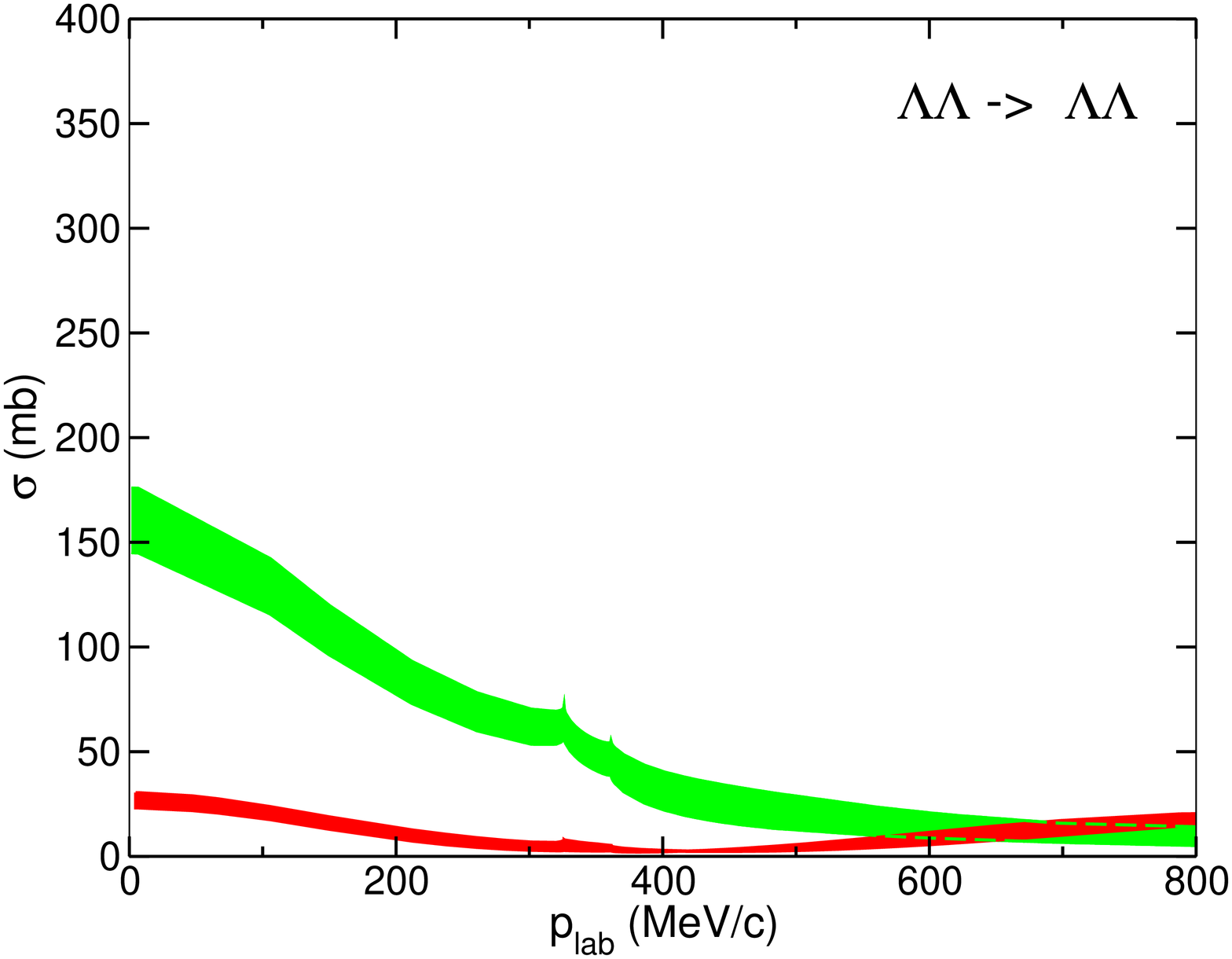}
\includegraphics*[width=6cm,angle=-90]{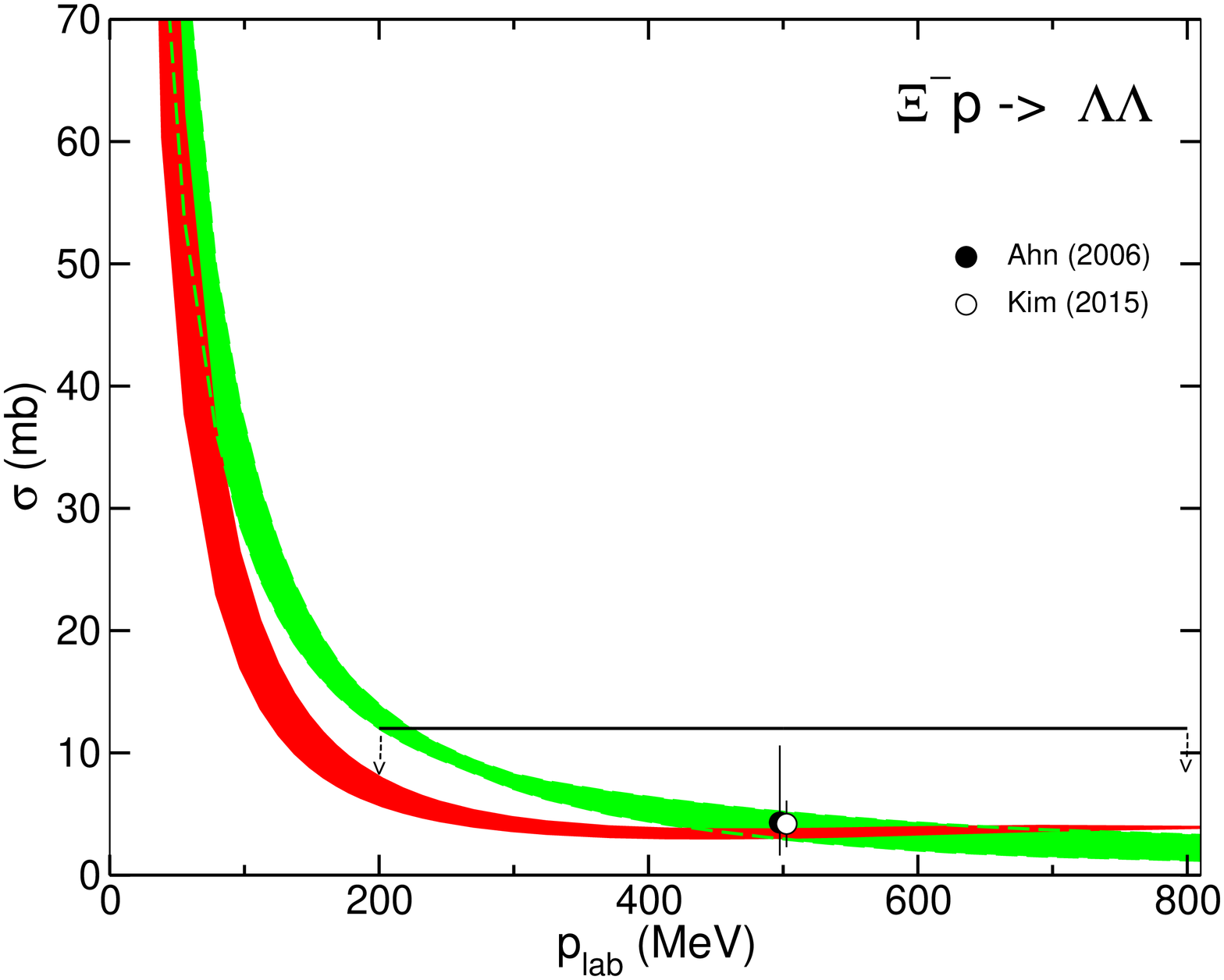}

\includegraphics*[width=6cm,angle=-90]{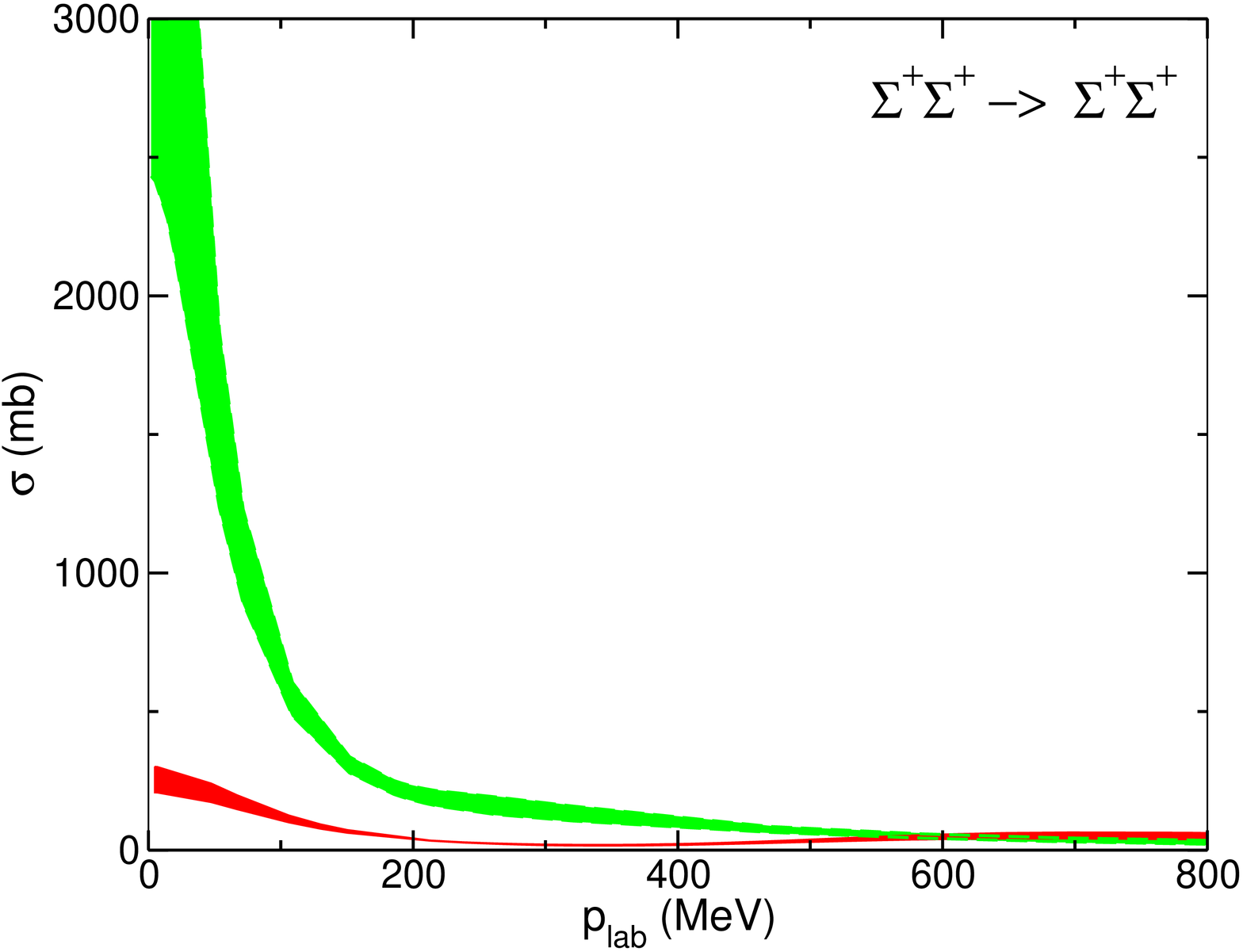}
\caption{$\Lambda\Lambda$ and $\Sigma^+\Sigma^+$ cross sections. 
The bands represent our results at NLO (black/red) and LO (grey/green).
The experimental information is taken from Ahn et al. \cite{Ahn:2006} (filled circle)
and Kim \cite{Kim:2015S} (open circle). Upper limits are indicated by arrows. 
}
\label{XS1}
\end{figure}

Let us start with the $\Lambda\Lambda$ and ($I=2$) $\Sigma\Sigma$ channels which are determined
solely by the symmetric SU(3) representations ($27$, $8_s$, $1$), see Table~\ref{tab:SU3}. 
Corresponding results are displayed in Fig.~\ref{XS1} (cross sections) and summarized in 
Table~\ref{ERE0} (effective range parameters). 
As discussed in the preceding section, we use here the LECs $\tilde C^{27}_{^1S_0}$ and 
$C^{27}_{^1S_0}$ determined in our combined analysis of the $pp$ and $\Sigma^+ p$ systems
\cite{HMP:2015} where SU(3) breaking in the leading term is incorporated. 
It is reassuring to see that the results based on those LECs are in agreement with the data 
points as well as with the upper limit for the reaction $\Xi^-p\to \Lambda\Lambda$ 
provided in Refs.~\cite{Ahn:2006,Kim:2015S}, see the black (red) band in Fig.~\ref{XS1}. 
In addition, the ${\Lambda\Lambda}$ $^1S_0$ scattering length is well within the range 
implied by studies of bound states and reactions involving the ${\Lambda\Lambda}$ interaction 
summarized in Sect.~II.A. 
For the LECs in the SU(3) singlet representation, $\tilde C^{1}_{^1S_0}$ and $C^{1}_{^1S_0}$,
values close to zero had to be chosen. Any somewhat larger positive values lead to 
scattering lengths smaller than $-0.5$~fm while negative values would result in a 
$H$-dibaryon like near-threshold bound state in the $\Xi N$ system. 
If one takes over the values for the $C^{27}$'s from the $YN$ fit \cite{Haidenbauer13}
then the ${\Lambda\Lambda}$  scattering length would be in the order of $-2$~fm, i.e. 
clearly too large in view of the aforementioned analyses. It turned out that it is not 
possible to achieve a reduction from $-2$ fm, say, to the preferred range of $-0.5$ to $-1$~fm 
by a variation of the LECs $\tilde C^{1}_{^1S_0}$ and $C^{1}_{^1S_0}$ within its natural range.
  
One can see from Fig.~\ref{XS1} that already the result at LO was consistent with the data 
for $\Xi^-p\to \Lambda\Lambda$ (grey/green band). On the other hand, the $\Lambda\Lambda$ 
scattering length might be somewhat too large based on the expectations discussed above. 

\begin{table}[t]
\caption{
Scattering lengths and effective ranges (in fm) for $\Lambda\Lambda$ and $\Sigma^+\Sigma^+$,
for various cut-off values (in MeV). 
}
\label{ERE0}
\vspace{0.2cm}
\centering
\renewcommand{\arraystretch}{1.3}
\begin{tabular}{|c|l|rrrr|rrrr|}
\hline
\multicolumn{2}{|c|}{}         &\multicolumn{4}{|c|}{NLO} & \multicolumn{4}{|c|}{LO} \\
\multicolumn{2}{|c|}{$\Lambda$}& $500$ &$550$ &$600$ &$650$ &$550$ &$600$ &$650$ &$700$\\
\hline
$\Lambda\Lambda$ &$a_{1S0}$ &$-0.62$ &$-0.61$ &$-0.66$ &$-0.70$ &$-1.52$ &$-1.52$ &$-1.54$ &$-1.67$ \\
                  &$r_{1S0}$ &$6.95$   &$6.06$   &$5.05$   &$4.56$   &$0.82$  &$0.59$  &$0.31$  &$0.34$  \\
\hline
\hline
${\Sigma^+\Sigma^+}$&$a_{1S0}$ &$-2.19$ &$-1.94$ &$-1.83$ &$-1.82$ &$-6.23$ &$-7.76$  &$-9.42$ &$-9.27$ \\
&$r_{1S0}$ &$5.67$  &$5.97$  &$6.05$  &$5.93$  &$2.17$  &$2.00$   &$1.88$  &$1.88$ \\
\hline
\end{tabular}
\renewcommand{\arraystretch}{1.0}
\end{table}

\begin{figure}[t]
\includegraphics*[width=6cm,angle=-90]{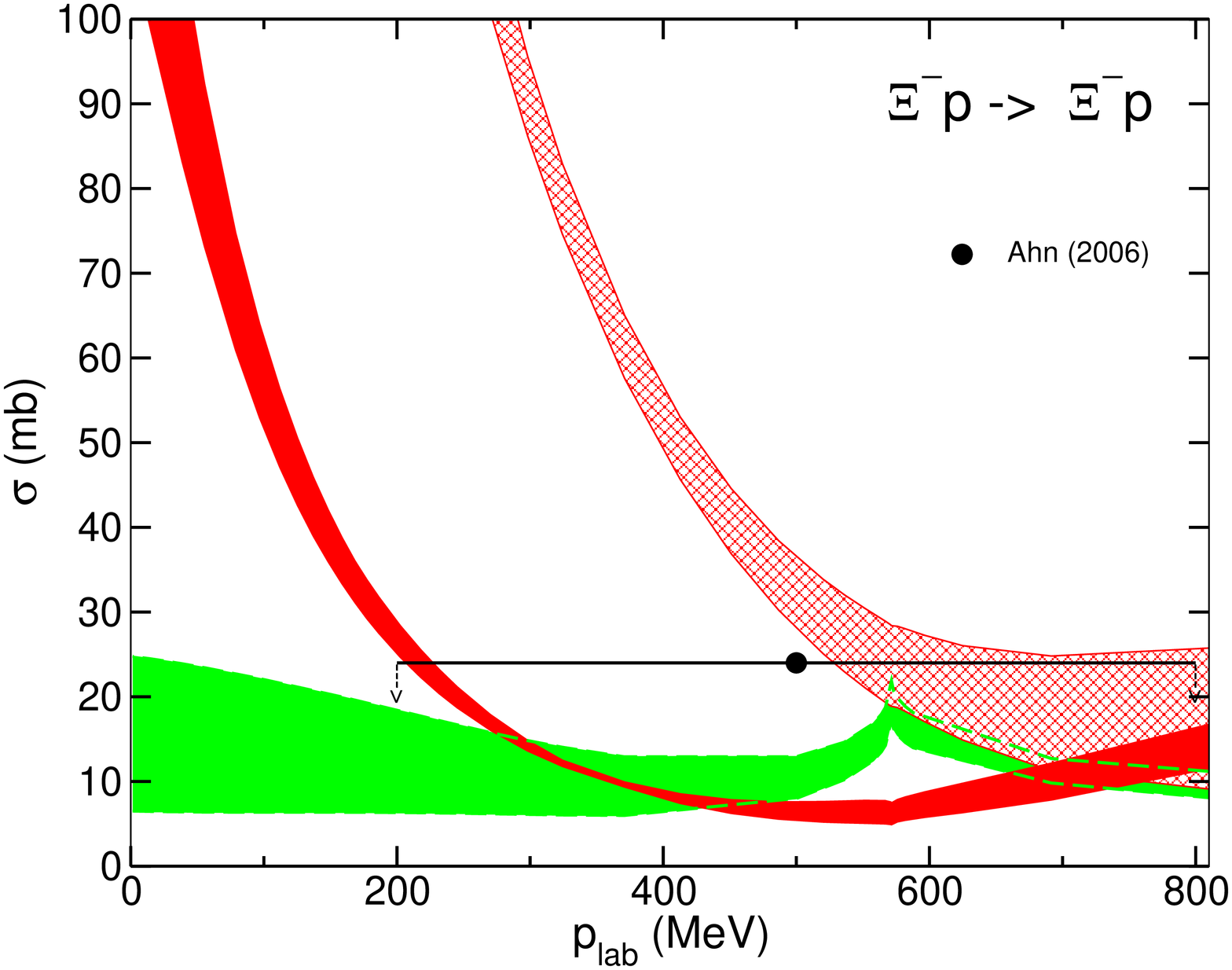}
\includegraphics*[width=6cm,angle=-90]{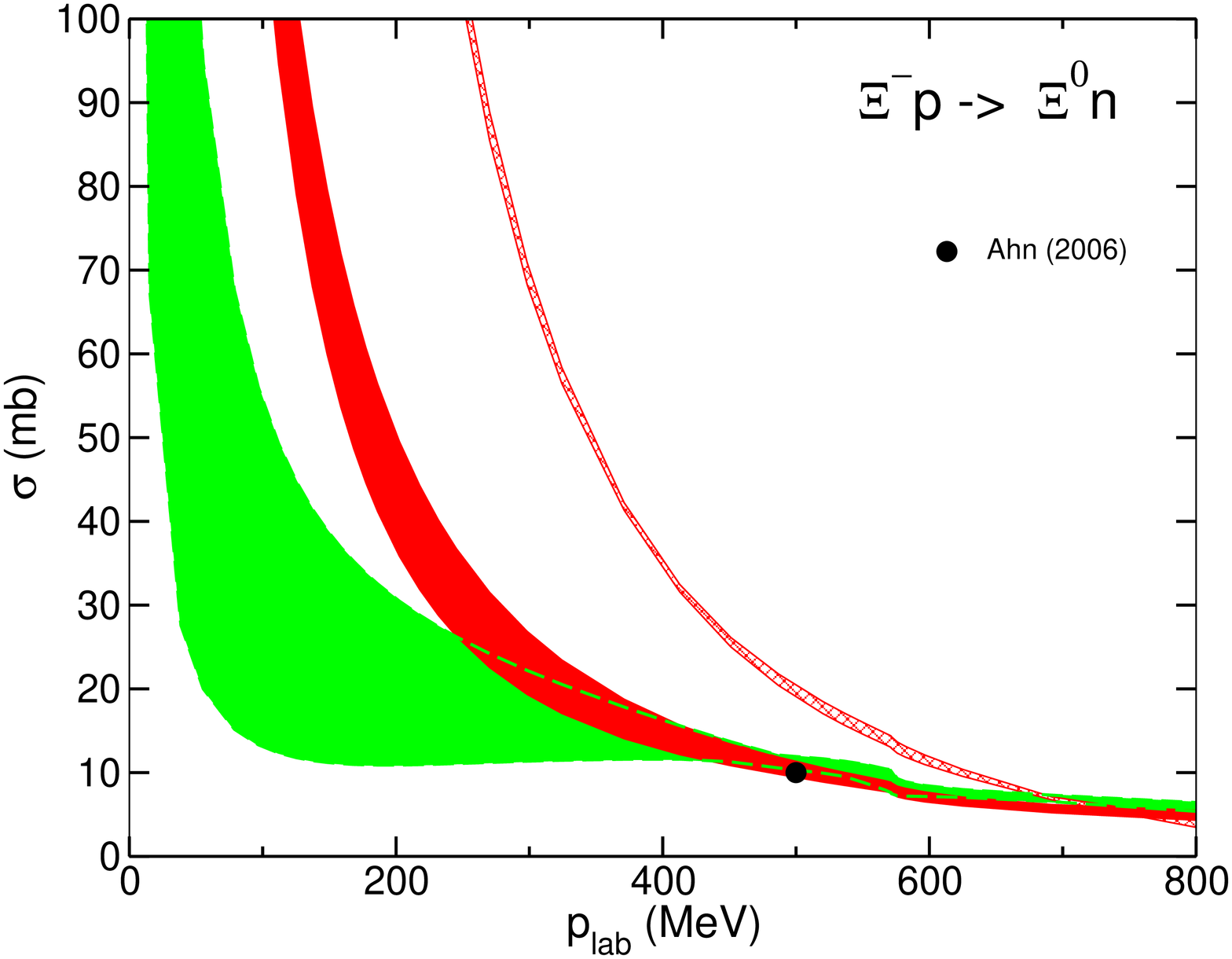}

\includegraphics*[width=6cm,angle=-90]{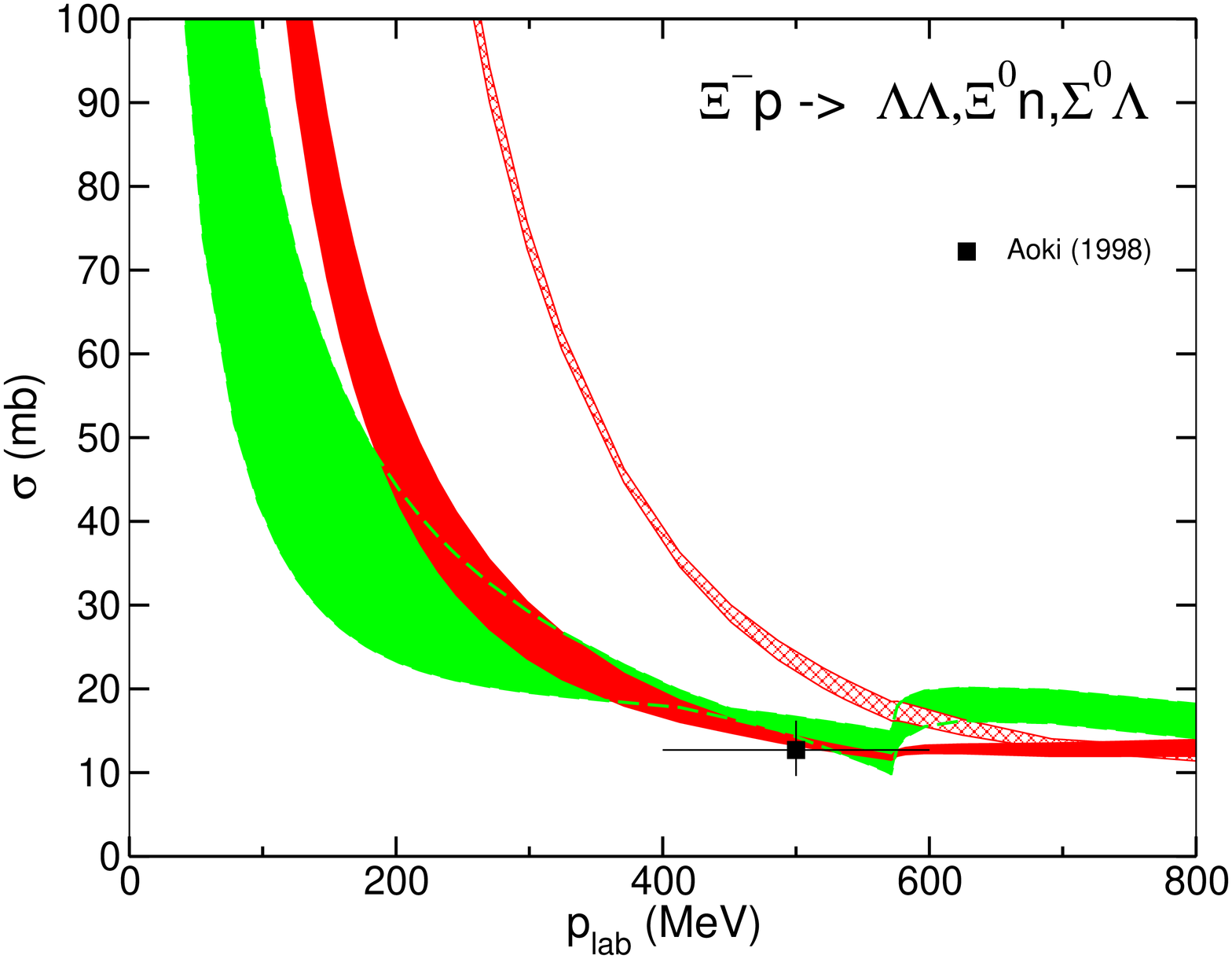}
\caption{$\Xi^- p$ induced cross sections. 
The bands represent our results at NLO (black/red) and LO (grey/green).
The hatched band shows results based on $\tilde C^{8_a}_{^3S_1}$ from
Ref.~\cite{Haidenbauer13}, see text. 
Experiments are from Ahn et al. \cite{Ahn:2006} and Aoki et al. \cite{Aoki:1998}.
Upper limits are indicated by arrows. 
}
\label{XS2}
\end{figure}

The $\Xi N$ interaction involves contributions from symmetric and antisymmetric
SU(3) representations, see Table~\ref{tab:SU3}. 
Results for $\Xi^-p$ elastic and inelastic cross sections and for the charge exchange 
reaction $\Xi^-p \to \Xi^0n$ are displayed in Fig.~\ref{XS2}. 
The black (red) bands represent those of our NLO interaction based on the re-adjusted 
leading contact term $\tilde C^{8_a}_{^3S_1}$, cf. the discussion in Sect.~III.B. 
Obviously, a nice agreement with the empirical information on the inelastic and charge-exchange 
cross sections is achieved and, moreover, the limits set by the measurement of Ahn et 
al.~\cite{Ahn:2006} for $\Xi^-p$ elastic scattering are also strictly fulfilled. 
Note that the inelastic cross section reported in Ref.~\cite{Aoki:1998} is actually 
for $\Xi^-N$. However, the first inelastic channel for $\Xi^-n$, namely $\Lambda\Sigma^-$, 
opens around $p_{\Xi^-}=590$~MeV/c in free scattering, i.e. practically above the
momentum range covered by the experiment. 
Therefore, it is sensible to compare this data point with our $\Xi^-p$ results.

For illustration purposes we display also results where all LECs in the 
antisymmetric SU(3) representations ($10^*$, $10$, $8_a$) were taken over from the
$YN$ fit \cite{Haidenbauer13}. Evidently, in this case the $\Xi^- p$ cross sections 
(hatched bands) are too large as compared to the experiments. 
Specifically, the charge-exchange as well as the inelastic cross section
are overestimated by roughly a factor two. Though the data in question come from 
in-medium measurements \cite{Aoki:1998,Ahn:2006} it is unrealistic to assume that
medium effects could explain the observed discrepancy. Indeed, at $\Xi$ 
momenta around $500$~MeV/c one would expect that such effects are moderate, say in the order 
of 10~\% or at most 20~\%, as suggested by corresponding calculations for the $NN$ system 
\cite{Li:1993,Schulze:1997,Kohno:1998}. 

Interestingly, the results based on our LO interaction from Ref.~\cite{Polinder07} 
(grey/green bands) are consistent with all empirical constraints. Those cross sections
are basically genuine predictions that follow from SU(3) symmetry utilizing LECs fixed from 
a fit to the $\Lambda N$ and $\Sigma N$ data on the LO level. 
The LO calculation exhibits also a
sizeable cusp effect in the $\Xi^-p$ cross sections at the opening of the $\Lambda\Sigma^0$ 
threshold which indicates a strong coupling to this channel. This effect is much smaller
for our NLO interaction and barely visible on the scale of Fig.~\ref{XS2}.

\begin{figure}[t]
\includegraphics*[width=6cm,angle=-90]{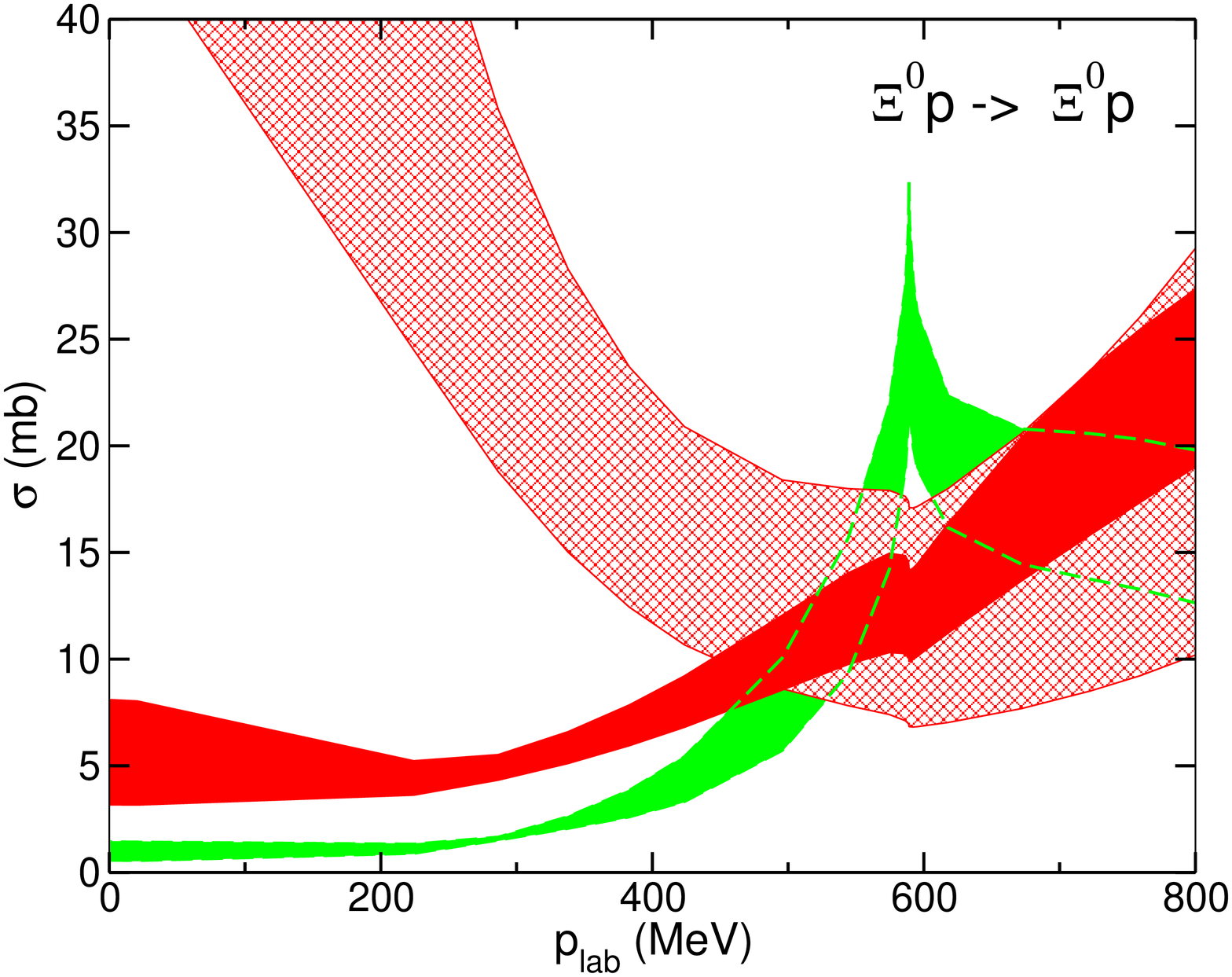}
\includegraphics*[width=6cm,angle=-90]{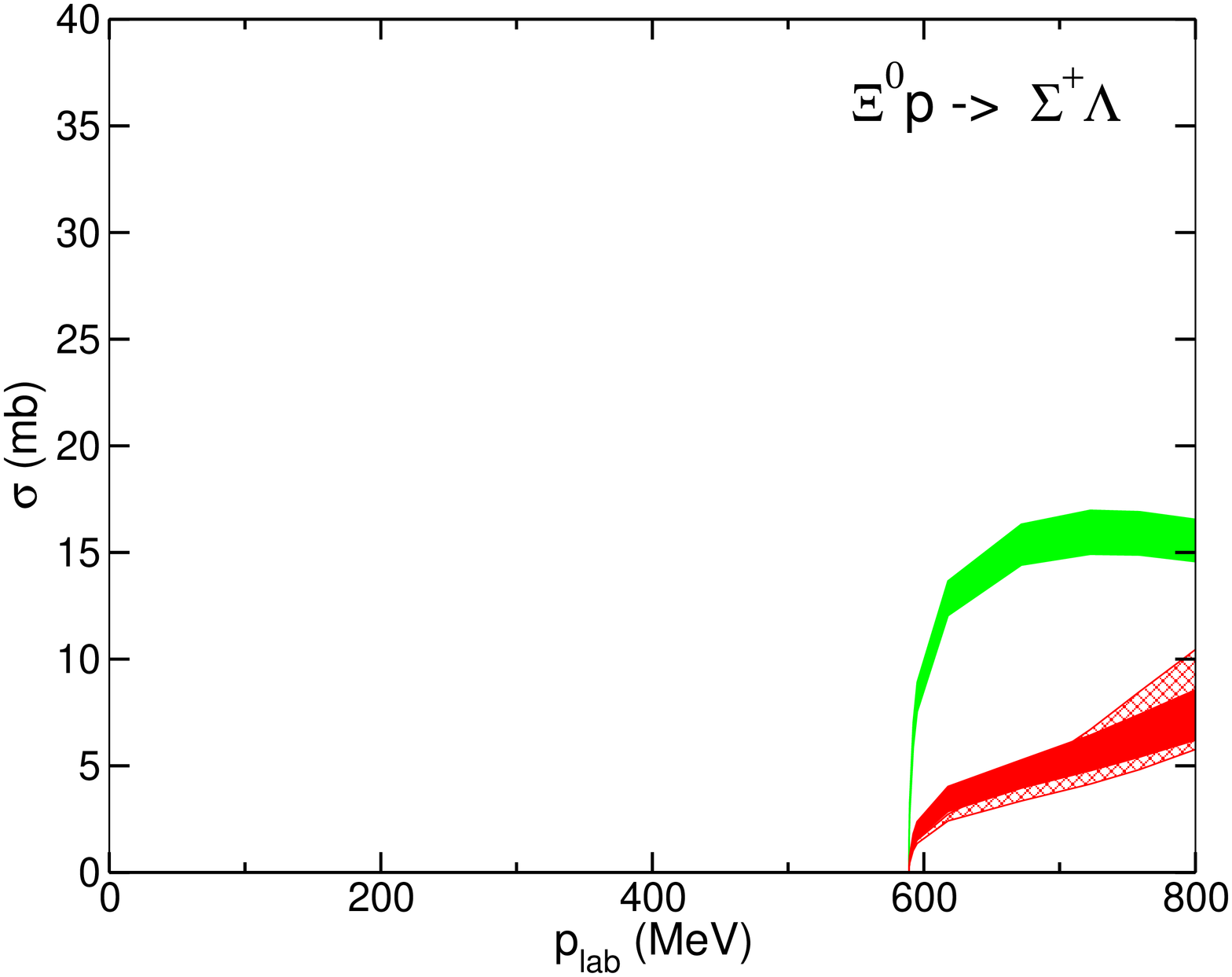}
\caption{$\Xi^0 p$ induced cross sections. 
The bands represent our results at NLO (black/red) and LO (grey/green).
The hatched band shows results based on $\tilde C^{8_a}_{^3S_1}$ from
Ref.~\cite{Haidenbauer13}, see text. 
}
\label{XS3}
\end{figure}

The $\Xi^0p$ system is a pure isospin $I=1$ state. 
Predictions for $\Xi^0p$ induced reactions are displayed in Fig.~\ref{XS3}. 
There are no data for the $\Xi^0p$ channel in the momentum region up to~$1$ GeV/c. 
The NLO and LO results for $\Xi^0p$ elastic scattering are of comparable order of magnitude.
Like before for $\Xi^-p$, employing the LEC $\tilde C^{8_a}_{^3S_1}$ from the $YN$ sector 
leads to a stronger interaction and would imply a much larger $\Xi^0p$ cross section near 
threshold (hatched band). Again the cusp effect (now at the $\Lambda\Sigma^+$ threshold) 
is strongly reduced for the NLO interaction. 
Data at higher momenta suggest a $\Xi^0p$ elastic cross section in the order of
$10$ to $30$~mb \cite{Muller,Charlton}. 
Tamagawa et al. have deduced the $\Xi^- N$ elastic cross section and also the ratio of
the $\Xi^- p$ to $\Xi^- n$ scattering cross section for an average $\Xi$ momentum 
of $550$~MeV/c \cite{Tamagawa}. $\Xi^- n$ is like $\Xi^0p$ a pure $I=1$ state so that
$\sigma_{\Xi^-n} \cong \sigma_{\Xi^0p}$. Thus, 
we conclude from Figs.~\ref{XS2} and \ref{XS3} that our results are compatible with 
their measurement. Specifically, we get 
$\sigma_{\Xi^-p}/\sigma_{\Xi^-n} \cong \sigma_{\Xi^-p}/\sigma_{\Xi^0p} \cong 1$
for $p_{\Xi^-} = 500\sim 600$~MeV/c as reported in Ref.~\cite{Tamagawa}.

The strong coupling between the $\Xi N$ and $\Lambda\Sigma$ systems at LO,
conjectured from the pronounced cusp in the $\Xi N$ cross section,
is indeed reflected in the cross section for $\Xi^0 p \to \Lambda\Sigma^+$.
It is significantly larger than the one predicted at NLO, see Fig.~\ref{XS3} (right side). 
A cross section of $24$~mb is given for this transition reaction in Ref.~\cite{Muller} 
for $\Xi$'s with an average momentum of $2$~GeV/c. 
The $\Sigma^0\Sigma^+$ channel opens at the $\Xi^0$ momentum of roughly $970$~MeV/c. 
A value of $6$~mb is given for the $\Xi^0 p \to \Sigma^0\Sigma^+$ transition
section in Ref.~\cite{Muller} for $\Xi$'s with an average momentum of $2$~GeV/c. 
We calculated this cross section for curiosity reasons for our NLO interaction and we 
found it to be in the order of $1$--$4$~mb for momenta around $1 \sim 1.3$~GeV/c. 

\begin{table}
\caption{Scattering lengths and effective ranges (in fm) for $\Xi^0p$ and $\Xi^0n$,
for various cut-off values (in MeV). Results for the isospin $I=0$ $\Xi N$ effective
range parameters, based on a calculation in isospin basis are also given. 
$(*)$ indicates the results based on $\tilde C^{8_a}_{^3S_1}$ from Ref.~\cite{Haidenbauer13}.
}
\label{ERE1}
\vspace{0.2cm}
\centering
\renewcommand{\arraystretch}{1.3}
\begin{tabular}{|c|l|rrrr|rrrr|}
\hline
\multicolumn{2}{|c|}{}         &\multicolumn{4}{|c|}{NLO} & \multicolumn{4}{|c|}{LO} \\
\multicolumn{2}{|c|}{$\Lambda$}& $500$ &$550$ &$600$ &$650$ &$550$ &$600$ &$650$ &$700$\\
\hline
${I=0}$&$a_{3S1}$         &$-0.33$ &$-0.39$ &$-0.62$ &$-0.85$    &$-0.85$ &$-0.59$ &$-0.43$ &$-0.32$ \\
&$r_{3S1}$         &$-6.87$  &$-1.77$  &$1.00$  &$1.42$   &$-1.84$ &$-3.93$ &$-7.16$ &$-12.56$  \\
&$a_{3S1}$ $(*)$   &$ 5.11$ &$ 4.03$ &$3.97$ &$3.97$ &        &        &        &        \\
&$r_{3S1}$ $(*)$   &$ 0.93$  &$ 1.11$  &$ 1.24$  &$ 1.27$  &        &        &        &        \\
\hline
\hline
${\Xi^0p}$&$a_{1S0}$         &$0.37$  &$0.39$  &$0.34$  &$0.31$  &$0.21$  &$0.19$        &$0.17$  &$0.13$ \\
&$r_{1S0}$         &$-4.71$ &$-4.86$ &$-7.07$ &$-8.99$ &$-30.7$ &$-37.7$       &$-52.8$ &$-98.5$ \\
&$a_{3S1}$         &$-0.20$ &$-0.04$ &$ 0.021$ &$ 0.039$ &$0.02$  &$0.00$        &$0.02$  &$0.03$ \\
&$r_{3S1}$         &$35.6$  &$575.1$  &$1797$  &$450$  &$968$   &$>\!\!10^4$   &$1166$  &$548$ \\
&$a_{3S1}$ $(*)$   &$-1.01$ &$-0.85$ &$-0.72$ &$-0.66$ &        &        &        &        \\
&$r_{3S1}$ $(*)$   &$ 3.60$ &$ 4.04$ &$ 4.60$  &$ 4.82$  &        &        &        &        \\
\hline
\hline
${\Xi^0n}$&$a_{3S1}$         &$-0.25$ &$-0.20$ &$-0.26$ &$-0.34$ &$-0.34$ &$-0.25$ &$-0.20$ &$-0.15$ \\
 &$r_{3S1}$         &$5.35$  &$8.36$  &$5.26$  &$2.93$  &$-5.86$ &$-8.27$ &$-4.36$ &$16.3$  \\
 &$a_{3S1}$ $(*)$   &$11.39$ &$ 5.15$ &$4.78$ &$4.74$ &        &        &        &        \\
 &$r_{3S1}$ $(*)$   &$-0.37$  &$-0.39$  &$-0.22$  &$-0.18$  &        &        &        &        \\
\hline
\end{tabular}
\renewcommand{\arraystretch}{1.0}
\end{table}
\begin{figure}[t]
\includegraphics*[width=5cm]{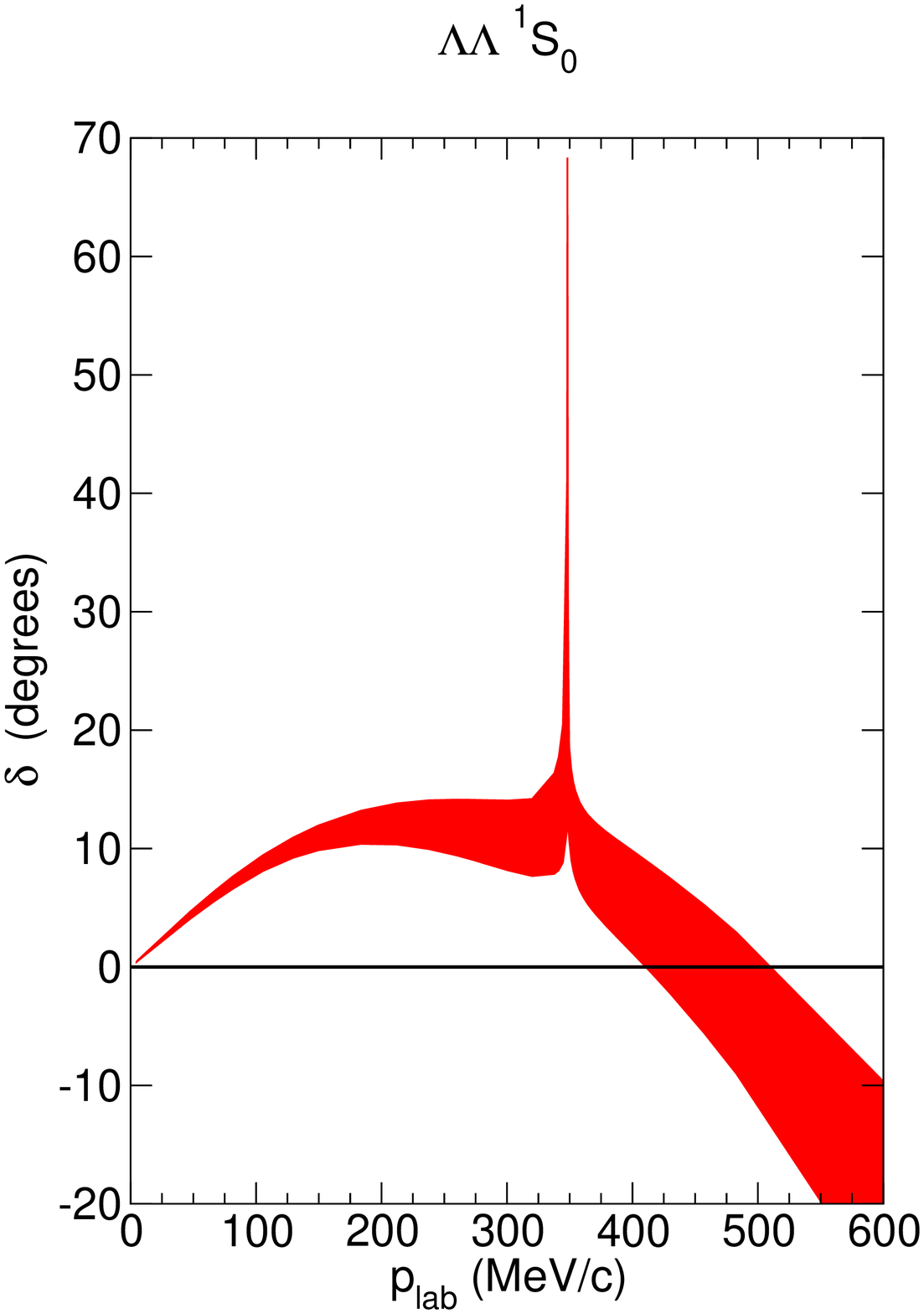}\includegraphics*[width=5cm]{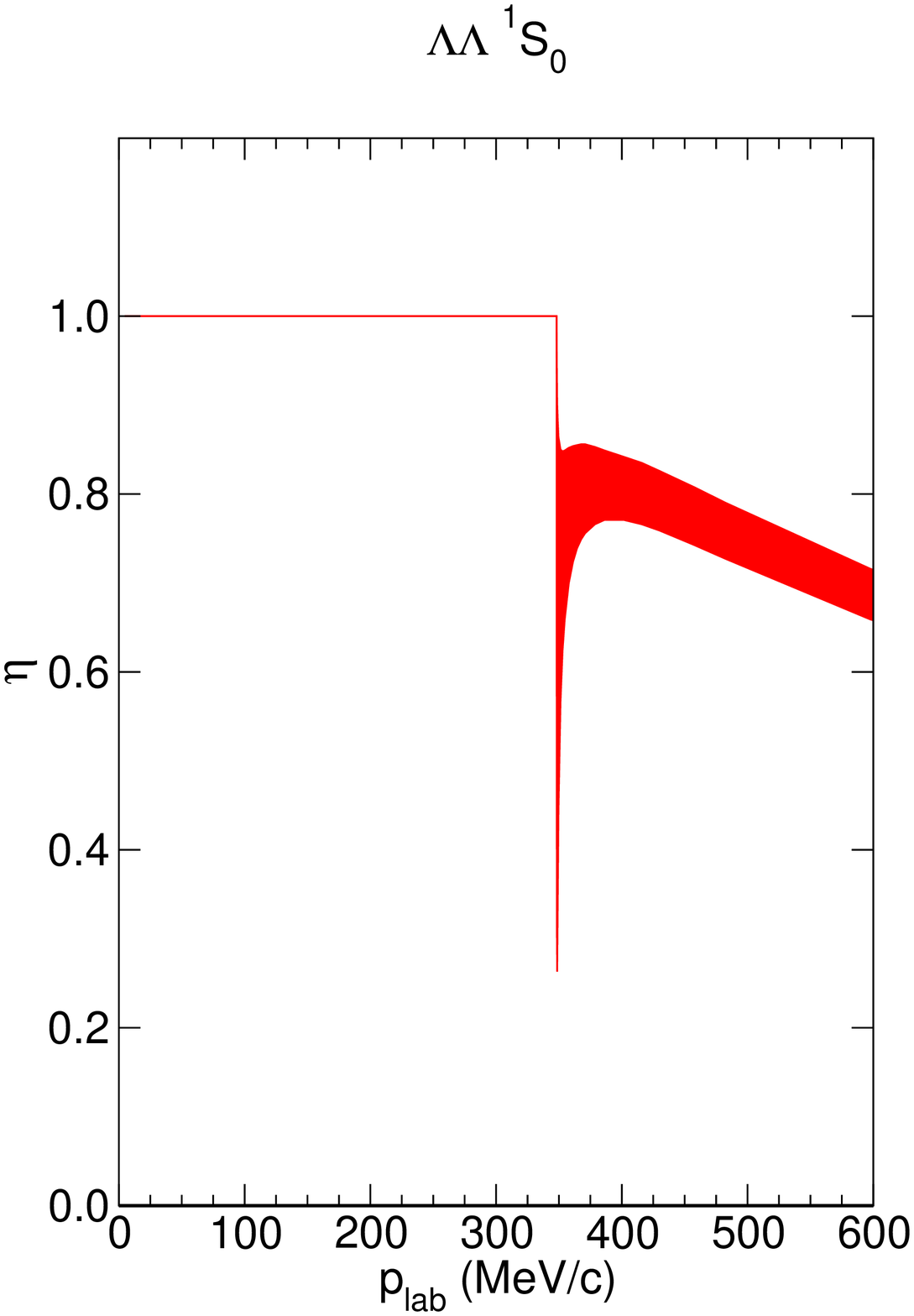}
\includegraphics*[width=5cm]{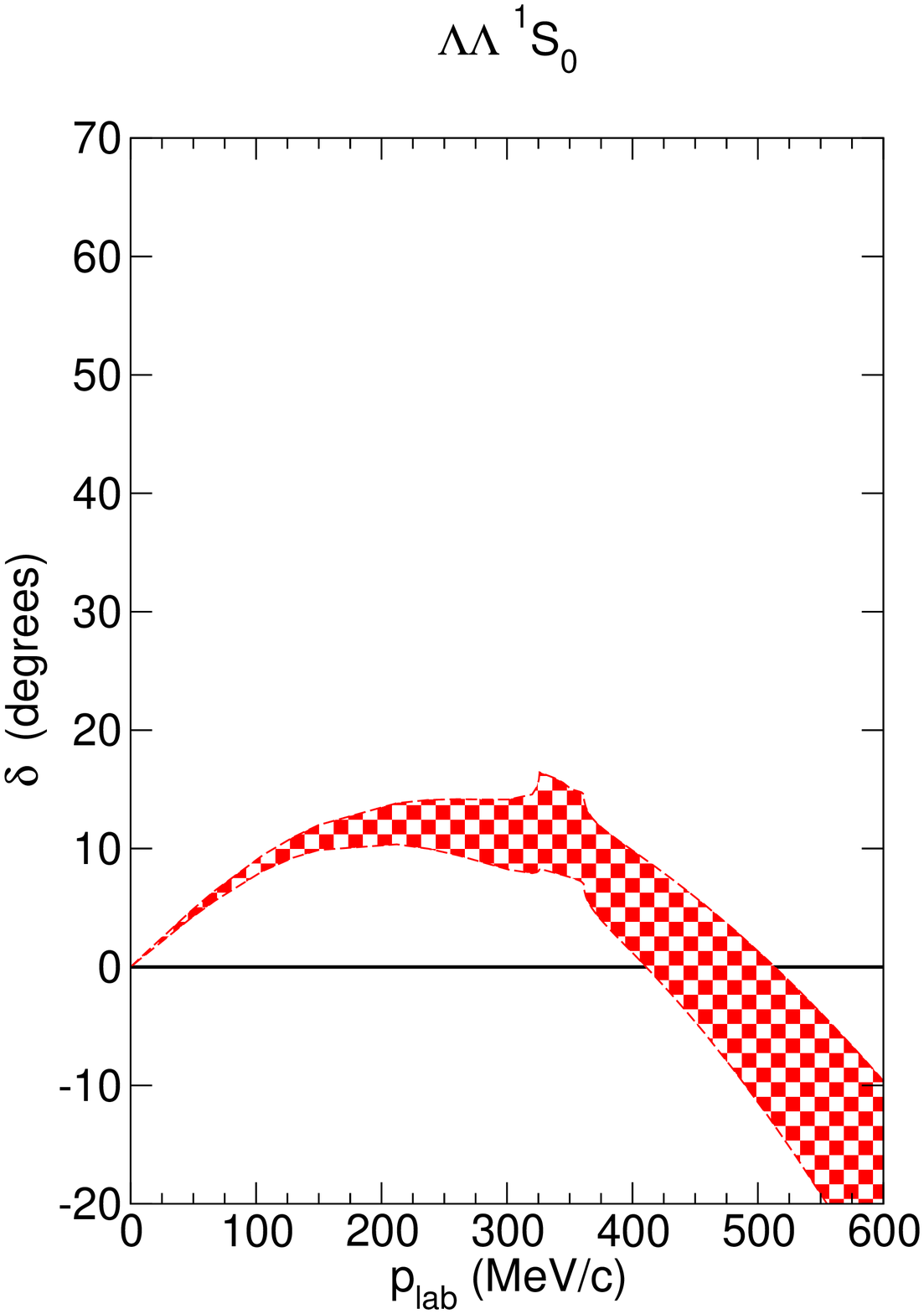}

\includegraphics*[width=5cm]{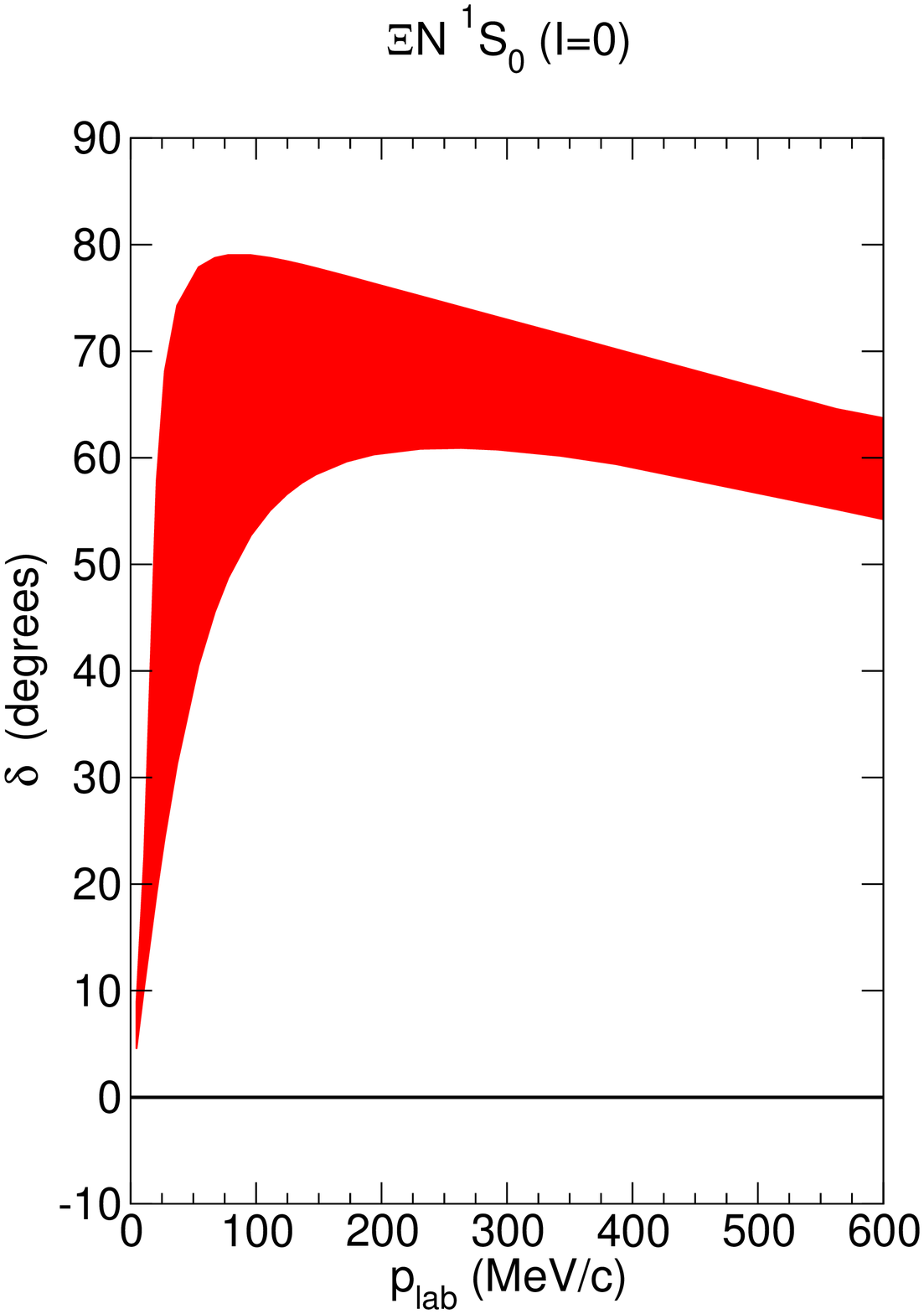}\includegraphics*[width=5cm]{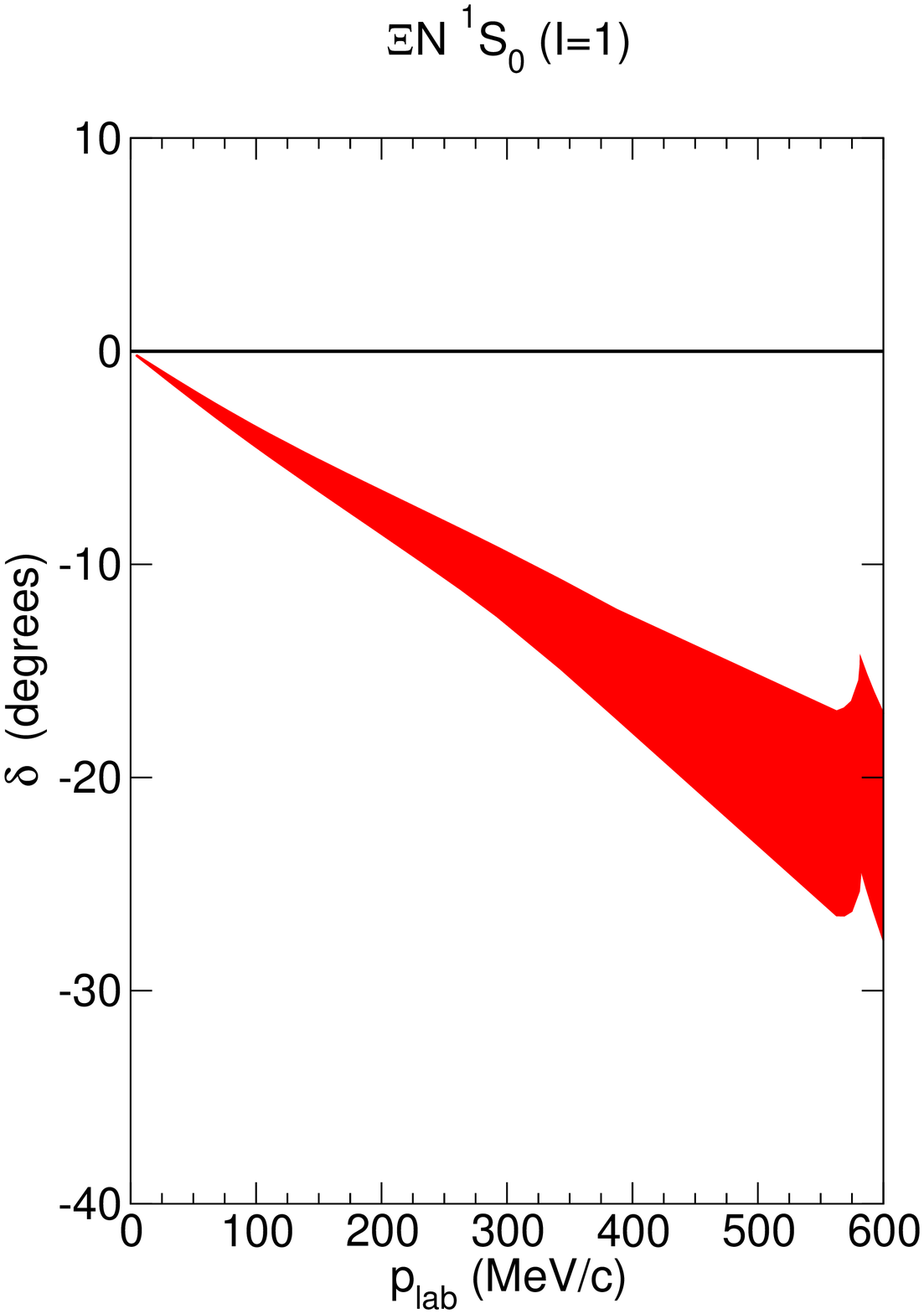}
\caption{$\Lambda\Lambda$ and $\Xi N$ phase shifts in the $^1S_0$ 
partial wave calculated in isospin basis. $\eta$ is defined by $S = \eta e^{2 i \delta}$.
The bands represent our results at NLO. 
The panel on the upper right side shows the result for $\Lambda\Lambda$ obtained in the particle basis.
}
\label{YYphase}
\end{figure}

\begin{figure}[t]
\includegraphics*[width=5cm]{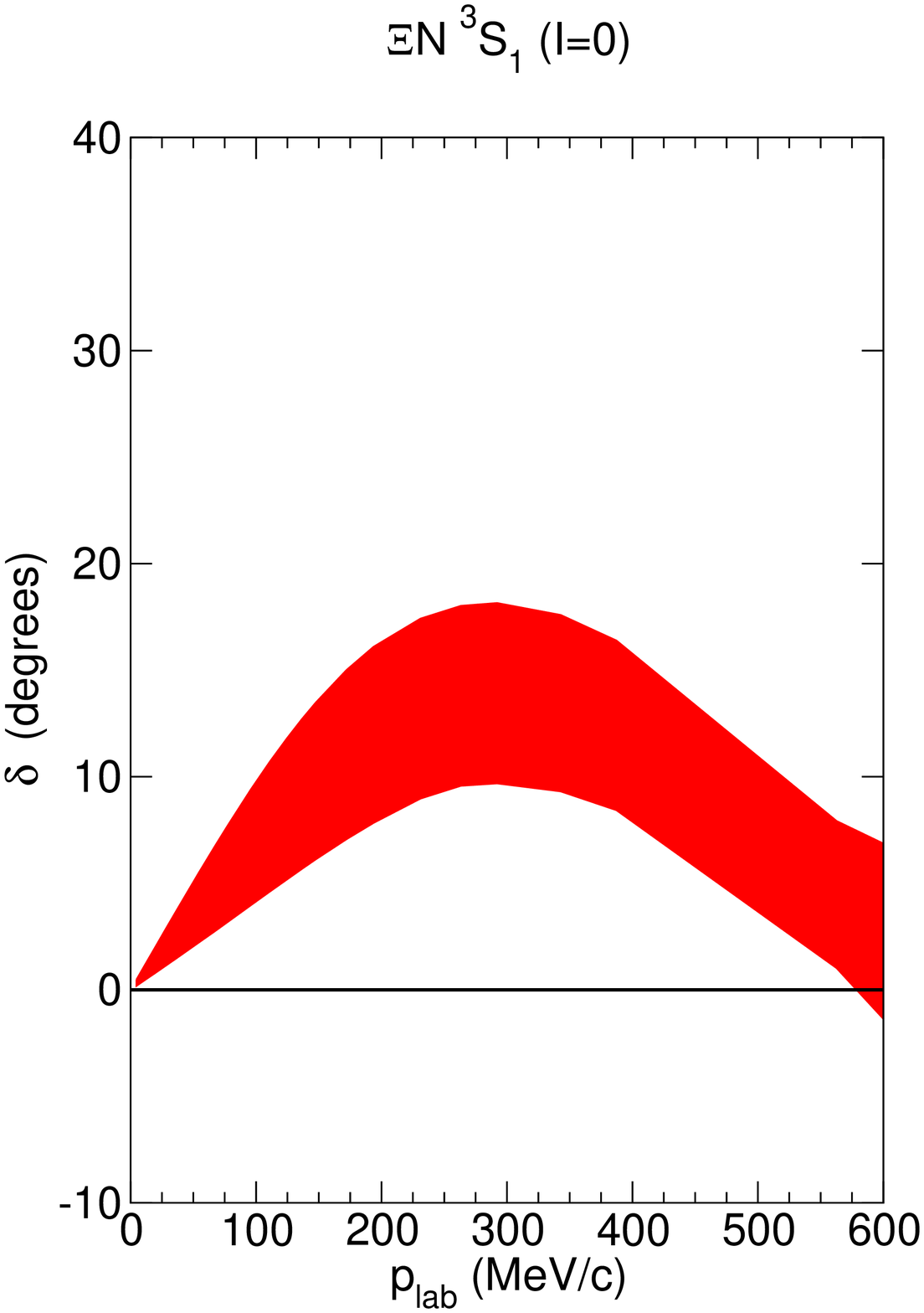}\includegraphics*[width=5cm]{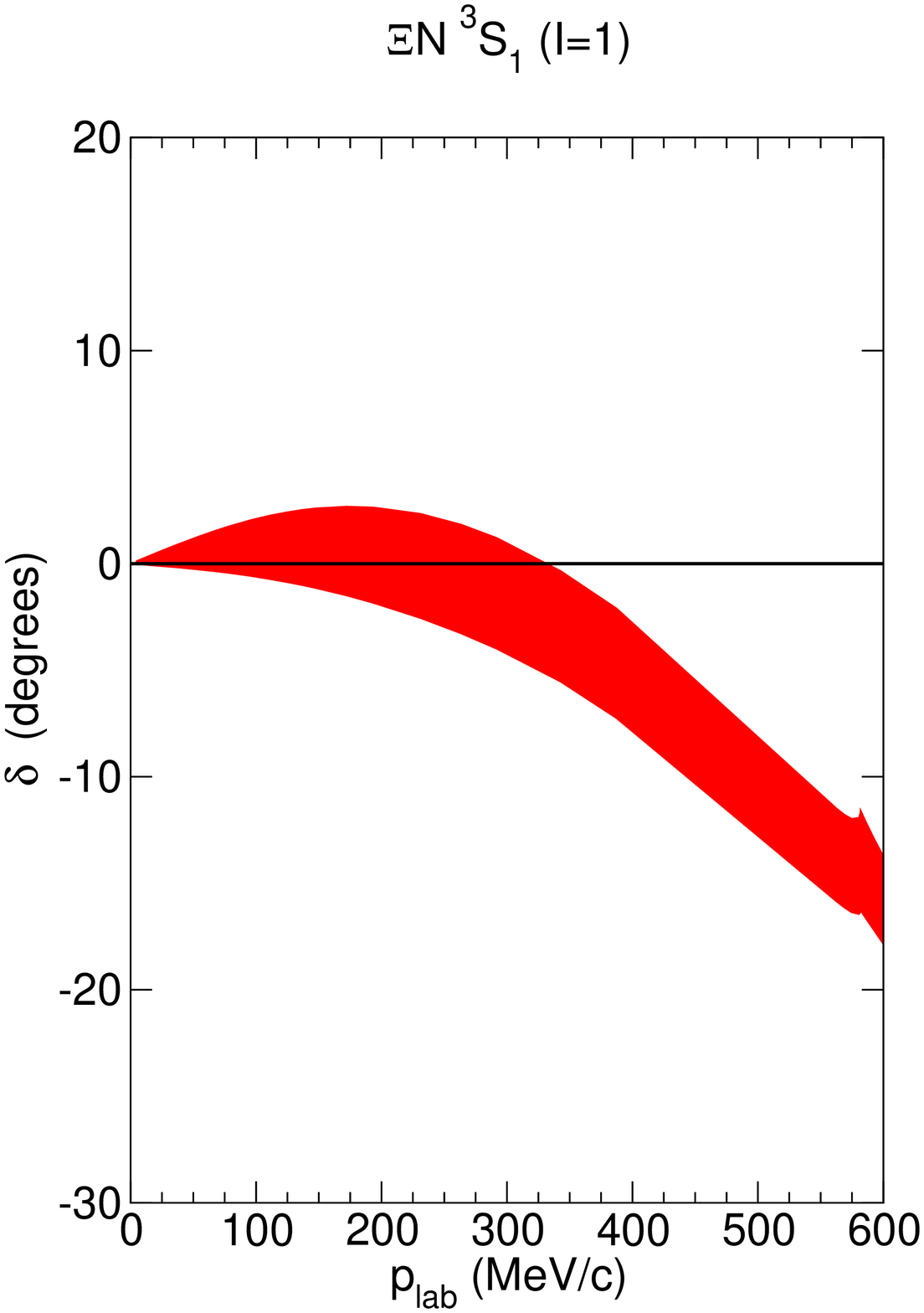}
\caption{$\Xi N$ phase shifts in the $^3S_1$ partial wave calculated in isospin basis. 
The bands represent our results at NLO. 
}
\label{YYphase2}
\end{figure}
In general, there is a noticeable reduction of the dependence of the $\Lambda\Lambda$ 
and $\Xi^- p$ results on the regularization in our NLO calculation as compared to LO.
The cut-off dependence (represented by the bands) provides a lower bound on the theoretical 
uncertainty and, accordingly, a rough estimate for the latter. 
It should be said, however, that one cannot expect that the bands resulting from the LO and the NLO 
interactions overlap for energies close to the threshold, as it is the case in investigations 
of the $NN$ system \cite{Epe05,Epelbaum:2014}. In contrast to $NN$ there 
are no data at small kinetic energies that pin down the scattering length quantitatively 
and, as a consequence, the predictions for the cross sections in that region. In fact, 
the situation for $S=-2$ is similar to the one in our study of the $YN$ system \cite{Haidenbauer13}  
where, for example, also the $\Lambda p$ cross sections at LO and NLO do not overlap close to threshold.
Note that recently improved schemes to estimate the theoretical error were 
proposed and applied to high-order $NN$ results \cite{Furnstahl2014,Epelbaum2015,Furnstahl2015}.
In addition, more efficient regularization schemes than the Gaussian regular function
employed in the present study have been advocated \cite{Epelbaum2015}. 

Table~\ref{ERE1} summarizes effective range parameters for the $\Xi^0 p$ and $\Xi^0 n$ 
channels. Furthermore, $\Xi N$ results for $I=0$ calculated in the isospin basis and 
using an average $\Xi$ mass of $1318.07$~MeV are provided. There is a large difference 
between the physical masses of the $\Xi^0$ and $\Xi^-$ \cite{PDG} which causes a noticeable 
isospin breaking in the $\Xi N$ system in the threshold region because the thresholds 
for the $\Xi^0 n$ and $\Xi^-p$ channels do not coincide (as they would in the isospin
symmetric case) but are separated by about $7$~MeV. 
On the other hand, the results for $I=1$ (based on averaged $\Xi$ and $N$ masses) are 
very close to those for $\Xi^0p$ and, therefore, we do not write them down separately. 

We list only effective range parameters for elastic channels. Thus, for isospin $I=0$ 
only results for the $^3S_1$ partial wave are given because in this case there is no 
coupling of $\Xi N$ to $\Lambda\Lambda$ due to the Pauli principle. 
One can see from Table~\ref{ERE1} that the NLO and LO potentials yield $I=0$ scattering lengths 
of comparable magnitude. In both cases a weakly attractive interaction is predicted. 
Also the cut-off dependence is comparable, though its effect is actually reversed at NLO as 
compared to the LO case. 
The NLO interaction based on the value of $\tilde C^{8_a}_{^3S_1}$ from Ref.~\cite{Haidenbauer13} 
is much more attractive, see the results marked with an asterisk ($*$) in Table~\ref{ERE1}. 
Indeed, the large and positive scattering length predicted here is in the order of 
the one in the $np$ $^3S_1$ partial wave where there is a bound state, namely the deuteron. 
Thus, connecting $YN$ and $\Xi N$ by strict SU(3) symmetry would imply the existence of a 
deuteron-like state in the $I=0$ $\Xi N$ system for our NLO interaction. 
However, as argued above, with such a strongly attractive force we are not able to meet the 
constraints set by the experiments. Hence, if one takes those constraints serious one 
would rather exclude the existence of such a state. 

The effective range parameters for $\Xi^0 p$ reveal that the interaction in the isospin $I=1$ 
channel is likewise weak. The small and positive values of the scattering length 
in the $^1S_0$ partial wave point to a weak repulsion. Again the results obtained 
at the NLO and LO level are comparable. With regard to the results for the $^3S_1$ partial wave 
based on $\tilde C^{8_a}_{^3S_1}$ from Ref.~\cite{Haidenbauer13} those are somewhat more 
attractive but there is no dramatic difference. In some cases, notably in the $^3S_1$ partial 
wave, we observe unnaturally large values for the effective range, which is clearly
related to the strong suppression of the corresponding scattering length. This
peculiarity occurs also in phenomenological models of the $BB$ 
interaction \cite{Stoks99,Fujiwara:2006,Gasparyan2011}.

When isospin symmetry is fulfilled then the $\Xi^0 n$ scattering length is given simply by 
$(a^{I=0}+a^{I=1})/2$. However, as mentioned above, the large splitting between the  
$\Xi^0$ and $\Xi^-$ masses causes a sizeable isospin breaking in the reaction amplitudes
when solving the coupled-channels LS equation (\ref{LS}) in the particle basis, 
see the actual results for $\Xi^0 n$ in Table~\ref{ERE1}. 
A striking isospin breaking effect occurs in the NLO interaction with $\tilde C^{8_a}_{^3S_1}$ 
from Ref.~\cite{Haidenbauer13} due to the presence of the near-threshold deuteron-like 
bound state. This state is now even closer to the threshold, as testified by the increase
in the scattering length as compared to the $I=0$ case for all cut-off values. 

Finally, for completeness, in Figs.~\ref{YYphase} and \ref{YYphase2}  we present phase shifts 
for the $\Lambda\Lambda$ and $\Xi N$ $^1S_0$ and $^3S_1$ partial waves for $I=0$ and $1$, 
calculated in the isospin basis employing isospin-averaged masses. 
In addition, the inelasticity parameter $\eta$ 
(based on the standard parametrization of the $S$-matrix, $S = \eta e^{2 i \delta}$) 
is shown for the coupled system $\Lambda\Lambda$--$\Xi N$--$\Si\Si$ in the $I=0$ channel. 
Results for the $\Sigma\Sigma$ channel with $I=2$ can be found in Fig.~\ref{NNphase}. 

The most interesting feature in the phase-shift results is certainly the spectacular
cusp that occurs in the $\Lambda\Lambda$ system at the opening of the $\Xi N$ channel.
It is the consequence of an inelastic virtual state \cite{Badalyan82} very close to the
$\Xi N$ threshold predicted by the NLO interaction. This state can be seen as a remnant 
of the $H$-dibaryon. Indeed, some extrapolations of the lattice QCD results indicate that 
the dibaryon could be actually located close to the $\Xi N$ threshold, see 
Refs.~\cite{JMH2,Inoue11a} and the $\Lambda\Lambda$ phase shifts presented in those works. 
A structure close to the $\Xi N$ threshold is also predicted by the quark-model
based $BB$ interaction FSS in Ref.~\cite{Fujiwara:2006}. 
Concering our results it should be said, however, that there is a drastical change 
when we perform the same calculation in particle basis. 
Then the $\Xi^0n$ and $\Xi^-p$ channels open at different energies and, in
particular, the virtual state is farther away from the lower threshold (the one of $\Xi^0n$).
Accordingly, in this case there are two cusps and both are rather modest, see the panel
on the right hand side of Fig.~\ref{YYphase}. This is also manifest in
the corresponding cross section (cf. Fig.~\ref{XS1}) where the opening of the 
$\Xi N$ channels is barely visible.
Cusps appear also in the $\Xi N$ $I=1$ phase shifts at the opening of the $\Sigma\Lambda$ 
channel, in the $^1S_0$ as well as in the $^3S_1$ partial waves.

For $I=0$ chiral EFT at NLO predicts positive values for both $S$-wave phase shifts, 
which signals an attractive $\Xi N$ force in that isospin channel, cf. also the discussion 
on the scattering lengths above. The phase shift in the $^1S_0$ partial wave is rather 
large as a result of the presence of the aforementioned virtual state.
In case of $I=1$ the phase shifts are predominantly negative and, accordingly, the
$\Xi N$ force essentially repulsive. It should be mentioned that qualitatively similar 
features have been reported in Ref.~\cite{Fujiwara:2006} for a $BB$ interaction 
derived in the quark model.

\section{Summary}

In this paper we presented a potential for the baryon-baryon interaction in the 
strangeness $S=-2$ sector, derived in chiral effective field theory to 
next-to-leading order in the Weinberg counting.
At the considered order there are contributions from one- and two-pseudoscalar-meson
exchange diagrams and from four-baryon contact terms without and with two
derivatives. As in case of our study to the $\La N$ and $\Si N$
systems \cite{Haidenbauer13}, SU(3) flavor symmetry is used as guiding principle 
in the derivation of the interaction. This means that all the coupling constants 
at the various baryon-baryon-meson and baryon-baryon-meson-meson
vertices are fixed from SU(3) symmetry and 
the symmetry is also exploited to derive relations between the contact terms.
Furthermore, contributions from all mesons of the pseudoscalar octet
($\pi$, $K$, $\eta$) are taken into account.
The SU(3) symmetry is, however, broken by the masses of the pseudoscalar
mesons and of the baryons for which we take the known physical values.

In the application of this scheme to the $S=-1$ sector we found that
one can achieve a combined description of the $\La N$ and $\Si N$
systems without any explicit SU(3) breaking in the contact interactions.
However, it also turned out that a simultaneous description of the $YN$ data 
and the $NN$ interaction, with contact terms fulfilling SU(3) symmetry 
strictly and consistently, is not possible. Specifically, the strength of 
the contact interaction in the $27$-representation that is needed to 
reproduce the $pp$ (or $np$)
$^1S_0$ phase shifts is simply not compatible with the one required
for the description of the empirical $\Si^+ p$ cross section \cite{Haidenbauer13}.
The same observation is now made in the extension to $S=-2$. Also here we
cannot simply take over all the low-energy constants as fixed from fitting
to the $\La N$ and $\Si N$ data. If we want to satisfy the limited and mostly
qualitative experimental constraints on the $\La \La$  and $\Xi N$ systems
then some of the LECs need to be re-adjusted. We want to emphasize 
that an SU(3) breaking in the leading $S$-wave contact terms is in line with the employed
power counting scheme \cite{Petschauer13} and with the results published in Ref.~\cite{HMP:2015}.

Our results suggest that the $\Xi N$ interaction has to be relatively 
weak in order to be in accordance with the available empirical constraints.
In particular, the published values and upper bounds for the $\Xi^- p$ elastic 
and inelastic cross sections \cite{Aoki:1998,Ahn:2006,Kim:2015S} practically rule out 
a somewhat stronger attractive $\Xi N$ force and, specifically, disfavors any near-threshold 
deuteron-like bound states in that system. However, it remains to be seen in how 
far such a weakly attractive $\Xi N$ interaction is in line with other experimental 
evidence such as a recently reported deeply bound $\Xi$ hypernucleus \cite{Nakazawa:2015}.
It should be also mentioned that the latest and still preliminary results from 
lattice QCD simulations apparently suggest a somewhat more attractive $\Xi N$ 
interaction \cite{Sasaki:2015}, notably in the $^3S_1$ partial wave with isospin 
$I=0$ \cite{Sasaki:2015S}. 
 
In any case, we want to emphasize that the present investigation, performed in 
chiral EFT up to NLO, should be considered to be of preliminary 
and exploratory nature. While in our preceding study for the $\La N$ and $\Si N$ 
interactions we were able to fix practically all pertinent contact terms (at least for 
the $S$ waves) by a direct fit to corresponding data, without any recourse to the $NN$ 
system, this is not possible for the strangeness $S=-2$ sector. Here we are 
not in a position that would allow us to determine the relevant LECs uniquely, 
due to the rather limited empirical information.
Planned experimental efforts at sites like J-PARC \cite{JPARC} in Japan and/or 
FAIR \cite{FAIR0,FAIR} in Darmstadt will hopefully lead to an appreciable improvement
of the data base in the not too far future and, thus, provide important 
further constraints on the strangeness $S=-2$ baryon-baryon interaction.    

\acknowledgments{
We thank Norbert Kaiser and Wolfram Weise for useful discussions. 
This work is supported in part by the DFG and the NSFC through funds provided to 
the Sino-German CRC 110 ``Symmetries and the Emergence of Structure in QCD''.
S.~Petschauer thanks the ``TUM Graduate School''.
Part of the numerical calculations has been performed on
the supercomputer cluster of the JSC, J\"ulich, Germany.
}

\appendix
\section{Coupling constants and isospin factors}
\label{app:A}
\countzero

For a baryon-baryon-meson interaction Lagrangian that is SU(3)-invariant 
the various coupling constants are related with each other by \cite{Swa63}
\begin{equation}
\begin{array}{rlrlrl}
f_{NN\pi}  = & f, & f_{NN\eta_8}  = & \frac{1}{\sqrt{3}}(4\alpha -1)f, & f_{\Lambda NK} = & -\frac{1}{\sqrt{3}}(1+2\alpha)f, \\
f_{\Xi\Xi\pi}  = & -(1-2\alpha)f, &  f_{\Xi\Xi\eta_8}  = & -\frac{1}{\sqrt{3}}(1+2\alpha )f, & f_{\Xi\Lambda K} = & \frac{1}{\sqrt{3}}(4\alpha-1)f, \\
f_{\Lambda\Sigma\pi}  = & \frac{2}{\sqrt{3}}(1-\alpha)f, & f_{\Sigma\Sigma\eta_8}  = & \frac{2}{\sqrt{3}}(1-\alpha )f, & f_{\Sigma NK} = & (1-2\alpha)f, \\
f_{\Sigma\Sigma\pi}  = & 2\alpha f, &  f_{\Lambda\Lambda\eta_8}  = & -\frac{2}{\sqrt{3}}(1-\alpha )f, & f_{\Xi\Sigma K} = & -f.
\end{array}
\label{su3}
\end{equation}
Thus, all coupling constants are given in terms of $f\equiv g_A/2f_0$ and the ratio $\alpha=F/(F+D)$.
We use the values $g_A= 1.26$ for the axial coupling constant and $f_0 \approx f_\pi = 93$~MeV for 
the pion decay constant, while for the so-called $F/(F+D)$-ratio we adopt 
the ${\rm SU(6)}$ value $\alpha=0.4$. The $\eta$ meson is identified with the
octet-state $\eta_8$.
In principle, there is an explicit SU(3) symmetry breaking in the
coupling constants as reflected in the different empirical values
of the decay constants $f_\pi$, $f_K$ and $f_\eta$ \cite{PDG}, which
we ignored in \cite{Haidenbauer13} and will do so also in the present work. 
The only SU(3) symmetry breaking in the meson-exchange contributions comes from
the masses of the pseudoscalar mesons $\pi$, $K$, and $\eta$,
for which we take the physical values \cite{PDG}. 

The one-pseudoscalar-meson-exchange potential is given by 
\begin{eqnarray}
V_{B_1B_2\to B_3B_4} 
&=&-f_{B_1B_3P}f_{B_2B_4P}\frac{\left(\mbox{\boldmath $\sigma$}_1\cdot{\bf q}\right)
\left(\mbox{\boldmath $\sigma$}_2\cdot{\bf q}\right)}{{\bf q}^2+m^2_P}\,{\mathcal I}_{B_1B_2\to B_3B_4}\ ,
\label{OBE}
\end{eqnarray}
where $f_{B_1B_3P}$, $f_{B_2B_4P}$ are the appropriate coupling constants as given in 
Eq.~(\ref{su3}) and $m_P$ is the actual mass of the exchanged pseudoscalar meson. 
The transferred momentum, ${\bf q}$, is definited in terms of the initial (${\bf p}$ ) and 
final (${\bf p}'$) center-of-mass (c.m.) momenta of the baryons as ${\bf q}={\bf p}'-{\bf p}$. 
The isospin factors ${\mathcal I}$ are summarized in Table~\ref{tab:OBE}.
Explicit expressions for the momentum- and spin-dependent part of the two-meson exchange potentials 
can be found in Appendix A of Ref.~\cite{Haidenbauer13}. 
Those expressions need to be multiplied with the appropriate combination of baryon-baryon-meson coupling 
constants and with the isospin factors.
Since the list of isospin factors for two-pseudoscalar-meson exchanges is rather bulky for the $S=-2$ sector
we collected the corresponding tables in a separate pdf file. 
The file can be obtained upon request directly from the authors.   
Note that the isospin factors for $\Si\Si$ with $I=2$ have been published in Ref.~\cite{HMP:2015}.

For the $\Xi N$ and $YY$ interactions there are couplings between 
channels with non-identical and with identical particles which requires special 
attention \cite{Miyagawa:1997ka}. We follow the treatment of the flavor-exchange 
potentials as done by the Nijmegen group. Then the proper anti-symmetrization of the states 
is achieved by multiplying specific transitions with $\sqrt{2}$ factors that are included 
in Table~\ref{tab:OBE}, see Refs.~\cite{Stoks99}. In Table~\ref{tab:OBE},
$P_f$ is the flavor-exchange operator having the values $P_f=1$ for even-L (spin) singlet and 
odd-L triplet partial waves (antisymmetric in spin-momentum space), and $P_f=-1$ for odd-L 
singlet and even-L triplet partial waves (symmetric in spin-momentum space). 
For simplification of the notation we also introduce the operators $P_+ \equiv (1+P_f)/2$ and 
$P_- \equiv (1-P_f)/2$. Then 
$P_+ = 1$ ($P_- = 0$) for even-L singlet and odd-L triplet partial waves
while 
$P_+ = 0$ ($P_- = 1$) for even-L triplet and odd-L single partial waves. 
We note that 
for $\Lambda\Lambda\rightarrow\Lambda\Lambda$, for example, $\eta$-exchange contributes 
only to spin-momentum space antisymmetric i.e. flavor symmetric partial waves, 
for example ${}^1S_0$, ${}^3P_{0,1,2}$, etc. 


\begin{table}
\caption{Isospin factors for the various one-pseudoscalar-meson exchanges. $P_f$ is the flavor-exchange operator, 
and $P_+ = (1+P_f)/2$, $P_- = (1-P_f)/2$. 
}
\label{tab:OBE}
\vskip 0.1cm
\renewcommand{\arraystretch}{1.2}
\centering
\begin{tabular}{|c|c|c c c|}
\hline
Channel &Isospin & \ $\pi$ \ & \ $K$ \ & \ $\eta$ \ \\
\hline
$\La\La\rightarrow \La\La$  &$0$    &$ 0$           &$0 $ &$P_+$ \\
$\La\La\rightarrow \Xi N $  &$0$    &$ 0$           &$2P_+ $ &$ 0 $ \\
$\La\La\rightarrow \Si\Si$  &$0$    &$-\sqrt{3}P_+$ &$0 $ &$ 0 $ \\
$\Xi N \rightarrow \Xi N $  &$0$    &$-3$ &$0 $ &$ 1 $ \\
$\Xi N \rightarrow \Si\Si$  &$0$    &$ 0$ &$2\sqrt{3}P_+$ &$ 0 $ \\
$\Si\Si\rightarrow \Si\Si$  &$0$    &$-2P_+$ &$0 $ &$P_+$ \\
\hline
$\Xi N \rightarrow \Xi N $  &$1$    &$ 1$ &$0 $ &$ 1 $ \\
$\Xi N \rightarrow \Si\La$  &$1$    &$ 0$ &$\sqrt{2} $ &$ 0 $ \\
$\Xi N \rightarrow \La\Si$  &$1$    &$ 0$ &$-\sqrt{2}P_f  $ &$ 0 $ \\
$\Xi N \rightarrow \Si\Si$  &$1$    &$ 0$ &$2\sqrt{2}P_-$ &$ 0 $ \\
$\Si\La\rightarrow \Si\La$  &$1$    &$ 0$ &$0 $ &$ 1 $ \\
$\Si\La\rightarrow \La\Si$  &$1$    &$P_f$ &$0 $ &$ 0 $ \\
$\Si\La\rightarrow \Si\Si$  &$1$    &$2P_- $ &$0 $ &$ 0 $ \\
$\Si\Si\rightarrow \Si\Si$  &$1$    &$-P_-$ &$0 $ &$P_-$ \\
\hline
$\Si\Si\rightarrow \Si\Si$  &$2$    &$P_+$ &$0$ &$P_+$ \\
\hline
\end{tabular}
\renewcommand{\arraystretch}{1.0}
\end{table}

For completeness all the LECs used in the present study are summarized in Tables~\ref{tab:F1} 
and \ref{tab:F2}. As said, most of the LECs are kept as given in Tables~3 and 4 of Ref.~\cite{Haidenbauer13}. 
In this context let us mention, however, that unfortunately
there are typos in Table 3 of that reference. The values for the 
LECs in the $10$ and $10^*$ representations should be interchanged (for the
$^3S_1$ partial wave and the $^3S_1$-$^3D_1$ transition). 
Furthermore, the values for $\tilde C^{10^*}_{^3S_1}$ (labelled as $\tilde C^{10}_{^3S_1}$
in Ref.~\cite{Haidenbauer13}, cf. above) should read 
$0.104$, $0.541$, $1.49$, $3.44$, $4.99$, and $5.60$ for increasing cut-offs, 
i.e. the comma was misplaced in case of the last 4 entries. 

\begin{table}
\caption{Contact terms for the $^1S_0$ and $^3S_1$-$^3D_1$
partial waves for various cut--offs. The values of the $\tilde C$'s are in
$10^4$ ${\rm GeV}^{-2}$ the ones of the $C$'s in $10^4$ ${\rm GeV}^{-4}$;
the values of $\Lambda$ in MeV.
$(*)$ indicates the value of $\tilde C^{8_a}_{^3S_1}$ from Ref.~\cite{Haidenbauer13}.
}
\renewcommand{\arraystretch}{1.2}
\label{tab:F1}
\vspace{0.2cm}
\centering
\begin{tabular}{|c|c|rrrr|}
\hline
\multicolumn{2}{|c|}{$\Lambda$}  &$500$ & $550$& $600$& $650$  \\
\hline
$^1S_0$
&$\tilde C^{27}_{^1S_0}$   &$ 0.1520$  &$0.3296$ &$0.6139$ &$1.0752$ \\
&$\tilde C^{8_s}_{^1S_0}$  &$0.1970$  &$0.1930$ &$0.1742$ &$0.1670$ \\
&$\tilde C^{1}_{^1S_0}$    &$-0.015$  &$-0.010$ &$ 0.000$ &$ 0.010$ \\
&$C^{27}_{^1S_0}$          &$2.260$  &$2.260$ &$2.260$ &$2.260$ \\
&$C^{8_s}_{^1S_0}$         &$-0.200$  &$-0.206$ &$-0.0816$ &$-0.0597$ \\
&$C^{1}_{^1S_0}$           &$ 0.000$  &$ 0.000$ &$ 0.000$ &$ 0.000$ \\
\hline
$^3S_1$-$^3D_1$
&$\tilde C^{10^*}_{^3S_1}$   &$0.541$  &$1.49$  &$3.44$  &$4.99$  \\
&$\tilde C^{10}_{^3S_1}$     &$0.209$  &$0.635$ &$1.420$ &$2.200$ \\ 
&$\tilde C^{8_a}_{^3S_1}$    &$0.070$  &$0.070$ &$0.080$ &$ 0.100$  \\
&$C^{10^*}_{^3S_1}$          &$2.310$  &$2.450$ &$2.740$ &$2.530$  \\
&$C^{10}_{^3S_1}$            &$0.143$  &$0.741$ &$1.090$ &$1.150$  \\
&$C^{8_a}_{^3S_1}$           &$0.469$  &$0.627$ &$0.775$ &$0.854$  \\
&$C^{10^*}_{^3S_1-\,^3D_1}$  &$-0.429$ &$-0.428$  &$-0.191$  &$-0.191$ \\ 
&$C^{10}_{^3S_1-\,^3D_1}$    &$-0.300$ &$-0.356$ &$-0.380$  &$-0.380$  \\
&$C^{8_a}_{^3S_1-\,^3D_1}$   &$0.0475$ &$0.0453$ &$-0.00621$ &$-0.00621$  \\
\hline
&$\tilde C^{8_a}_{^3S_1}$ $(*)$ &$0.00715$ &$-0.0143$ &$-0.0276$ &$-0.0269$\\
\hline
\end{tabular}
\renewcommand{\arraystretch}{1.0}
\end{table}
\begin{table}
\caption{Contact terms for the $P$-waves for various cut--offs.
The values of the LECs are in
$10^4$ ${\rm GeV}^{-4}$; the values of $\Lambda$ in MeV.
}
\renewcommand{\arraystretch}{1.2}
\label{tab:F2}
\vspace{0.2cm}
\centering
\begin{tabular}{|c|c|rrrr|}
\hline
\multicolumn{2}{|c|}{$\Lambda$} & $500$ & $550$& $600$& $650$ \\
\hline
$^3P_0$
&$C^{27}_{^3P_0}$  &$1.49$  &$1.51$ &$1.55$ &$1.60$ \\
&$C^{8_s}_{^3P_0}$ &$2.50$  &$2.50$ &$2.50$ &$2.50$ \\
&$C^{1}_{^3P_0}$   &$-0.30$  &$-0.30$ &$-0.30$ &$-0.30$ \\
$^3P_1$
&$C^{27}_{^3P_1}$  &$-0.43$  &$-0.43$ &$-0.43$ &$-0.43$ \\
&$C^{8_s}_{^3P_1}$ &$0.65$  &$0.65$ &$0.65$ &$0.65$ \\
&$C^{1}_{^3P_1}$   &$-0.30$  &$-0.30$ &$-0.30$ &$-0.30$ \\
$^3P_2$
&$C^{27}_{^3P_2}$  &$-0.063$ &$-0.041$ &$-0.025$ &$-0.012$ \\
&$C^{8_s}_{^3P_2}$ &$1.00$   &$1.00$ &$1.00$ &$1.00$ \\
&$C^{1}_{^3P_2}$   &$-0.30$  &$-0.30$ &$-0.30$ &$-0.30$ \\
\hline
$^1P_1$
&$C^{10}_{^1P_1}$   &$0.49$  &$0.49$ &$0.49$  &$0.49$  \\
&$C^{10^*}_{^1P_1}$ &$-0.14$  &$-0.14$ &$-0.14$  &$-0.14$\\  
&$C^{8_a}_{^1P_1}$  &$-0.35$  &$-0.35$ &$-0.35$  &$-0.35$  \\
\hline
$^1P_1$-$^3P_1$
&$C^{8_s8_a}_{^1P_1-\,^3P_1}$ &$ 0$  &$ 0$ &$ 0$  &$0$  \\
\hline
\end{tabular}
\renewcommand{\arraystretch}{1.0}
\end{table}

\end{document}